\documentclass[pdflatex, sn-mathphys-num, referee]{sn-jnl}

\usepackage{graphicx}%
\usepackage{lmodern}%
\usepackage{multirow}%
\usepackage{siunitx,upgreek}%
\usepackage{amsmath,amssymb,amsfonts}%
\usepackage{amsthm}%
\usepackage{mathrsfs}%
\usepackage[title]{appendix}%
\usepackage{textcomp}%
\usepackage{manyfoot}%
\usepackage{booktabs}%
\usepackage{algorithm}%
\usepackage{algorithmicx}%
\usepackage{algpseudocode}%
\usepackage{listings}%
\usepackage{booktabs}
\usepackage[table,xcdraw]{xcolor}
\usepackage{fancybox}
\usepackage{cancel}

\unnumbered

\begin{document}
\title[Article Title]{Integrated sub-terahertz cavity electro-optic transduction}

\author*[1,3,4]{\fnm{Kevin K. S.} \sur{Multani}}\email{kmultani@stanford.edu}
\equalcont{These authors contributed equally to this work.}

\author*[2,4]{\fnm{Jason F.} \sur{Herrmann}}
\email{jfherrm@stanford.edu}
\equalcont{These authors contributed equally to this work.}

\author[3]{\fnm{Emilio A.} \sur{Nanni}}

\author*[2,4]{\fnm{Amir H.} \sur{Safavi-Naeini}}\email{safavi@stanford.edu}

\affil[1]{\orgdiv{Department of Physics}, \orgname{Stanford University}, \orgaddress{\city{Stanford}, \country{USA}}}
\affil[2]{\orgdiv{Department of Applied Physics}, \orgname{Stanford University}, \city{Stanford}, \country{USA}}
\affil[3]{\orgdiv{SLAC National Accelerator Laboratory}, \orgname{Stanford University}, \city{Menlo Park}, \country{USA}}
\affil[4]{\orgdiv{E.L. Ginzton Laboratory}, \orgname{Stanford University}, \orgaddress{\city{Stanford}, \country{USA}}}

%%==================================%%
%%           ABSTRACT               %%
%%==================================%%

%TC:ignore
\abstract{
Emerging communications and computing technologies will rely ever-more on expanding the useful radio frequency (RF) spectrum into the sub-THz and THz frequency range. Both classical and quantum applications would benefit from advancing integration and incorporation of sub-THz and electro-optic technologies into common devices, such as modulators. Here we demonstrate an integrated triply-resonant, superconducting electro-optic transducer. Our design incorporates an on-chip $107$ GHz sub-THz niobium titanium nitride superconducting resonator, modulating a thin-film lithium niobate optical racetrack resonator operating at telecom wavelengths. We observe a maximum photon transduction efficiency of $\eta_{\text{OE}}\approx 0.82\times 10^{-6}$ and an average single-photon electro-optic interaction rate of $g_0/2\pi\approx\qty{0.7}{\kHz}$. We also present a study and analysis of the challenges associated with the design of integrated sub-THz resonators and propose possible solutions to these challenges. Our work paves the way for further advancements in resonant electro-optic technologies operating at sub-THz frequencies.
}
%TC:endignore
\maketitle

%%==================================%%
%%          INTRODUCTION            %%
%%==================================%%
\section*{Introduction}\label{introduction}
%%%%%%%%%%%

Next-generation communications, imaging, and sensing technologies rely on harnessing sub-terahertz (sub-THz) and terahertz (THz) frequencies. Occupying a spectral region between electronics and optics, the sub-THz (\qtyrange{0.09}{0.3}{\THz}, \qtyrange{3000}{1000}{\um}) and THz (\qtyrange{0.1}{10}{\THz}, \qtyrange{3000}{30}{\um}) frequencies offer advantages such as increased data transmission rates~\cite{rajabali:2023:presentfuture,headland:2023:terahertzintegration}, improved imaging resolution~\cite{khalatpour:2021:highpowerportable, li:2024:plasmonicphotoconductive}, new radar and ranging modalities~\cite{zhu:2025:integratedlithium}, and access to new regimes of physical phenomena~\cite{bera:2021:reviewrecent}. Sub-THz frequencies offer several compelling advantages for quantum science applications beyond conventional microwave systems. Operating at elevated frequencies allows superconducting processors to function at higher temperatures with access to greater cooling power~\cite{anferov:2024:superconductingqubits}, potentially reducing the scaling barriers for fault-tolerant quantum computing~\cite{multani:2024:quantumlimits1}. Building on these developments, there is a nascent effort toward creating quantum devices in the millimeter-wave range (30-\qty{300}{\GHz}, 10 - \qty{0.1}{\mm}), as demonstrated by recent breakthroughs in millimeter-wave (mm-wave) superconducting qubits~\cite{anferov:2024:improvedcoherence, anferov:2024:millimeterwavesuperconducting}. This emerging quantum toolbox includes innovations such as neutral atom systems serving as quantum-enabled millimeter-wave-to-optical transducers with high conversion efficiency~\cite{kumar:2023:quantumenabledmillimetre}, sub-THz electromechanical ~\cite{xie:2023:subterahertzelectromechanics12} and optomechanical systems~\cite{hauer:2021:quantumoptomechanics,xie:2024:subterahertzoptomechanics}, and mm-wave phononic bandgap structures that couple to defects in diamond~\cite{kuruma:2025:controllinginteractions}. Important missing elements are integrated transduction mechanisms between millimeter-wave and photonic systems essential for networking these devices and enabling hybrid quantum systems.

Frequency conversion and transmission of information between disparate systems are essential in classical and quantum information systems and sensors. Although classical signals can be addressed with commercially available modulators, quantum signals require specialized transducers. Quantum transducers, with their significantly different figures of merit and requirements, have only recently garnered significant attention. During the past decade, integrated quantum transducers interconverting between microwave (\qty{3}{\GHz}) and optical (\qty{193.5}{\THz}) frequencies have been developed using direct electro-optic~\cite{mckenna:2020:cryogenicmicrowavetooptical, holzgrafe:2020:cavityelectrooptics, benea-chelmus:2020:electroopticinterface, tsang:2010:cavityquantum, tsang:2011:cavityquantum}, and piezo-, electro-, and opto-mechanical approaches~\cite{jiang:2023:opticallyheralded, meesala:2024:nonclassicalmicrowave, mirhosseini:2020:superconductingqubit,zhao:2023:electrooptictransduction,bozkurt:2023:quantumelectromechanical}.

In this work, we present a triply-resonant sub-THz cavity electro-optical platform based on thin-film lithium niobate (TFLN). We provide details regarding the superconducting sub-THz resonator and TFLN optical cavity, presenting crucial analysis needed to achieve the quantum operation of such sub-THz devices. Leveraging integrated photonics and RF co-packaging, our device is capable of sub-THz-to-telecom direct electro-optic transduction. A fully fledged device could link systems operating at elevated frequencies and temperatures~\cite{multani:2024:quantumlimits1, anferov:2024:superconductingqubits}. Examples of such systems include superconducting qubits~\cite{anferov:2024:millimeterwavesuperconducting}, neutral atoms~\cite{kumar:2023:quantumenabledmillimetre}, and long-baseline telescopes~\cite{gottesman:2012:longerbaselinetelescopes, wootten:2009:atacamalarge}. Additionally, such a transducer may also be used in multistage transduction to link current quantum hardware at microwave frequencies to optical frequencies, via a sub-THz intermediary~\cite{pechal:2017:millimeterwaveinterconnects, wang:2022:generalizedmatching, sahbaz:2024:terahertzmediatedmicrowavetooptical}. 

%%==================================%%
%%            RESULTS               %%
%%==================================%%

\section*{Results}
\label{results}
\subsection*{Device overview \& operating principle}
\begin{figure*}[]
    \centering
    \includegraphics[width=1.0\linewidth]{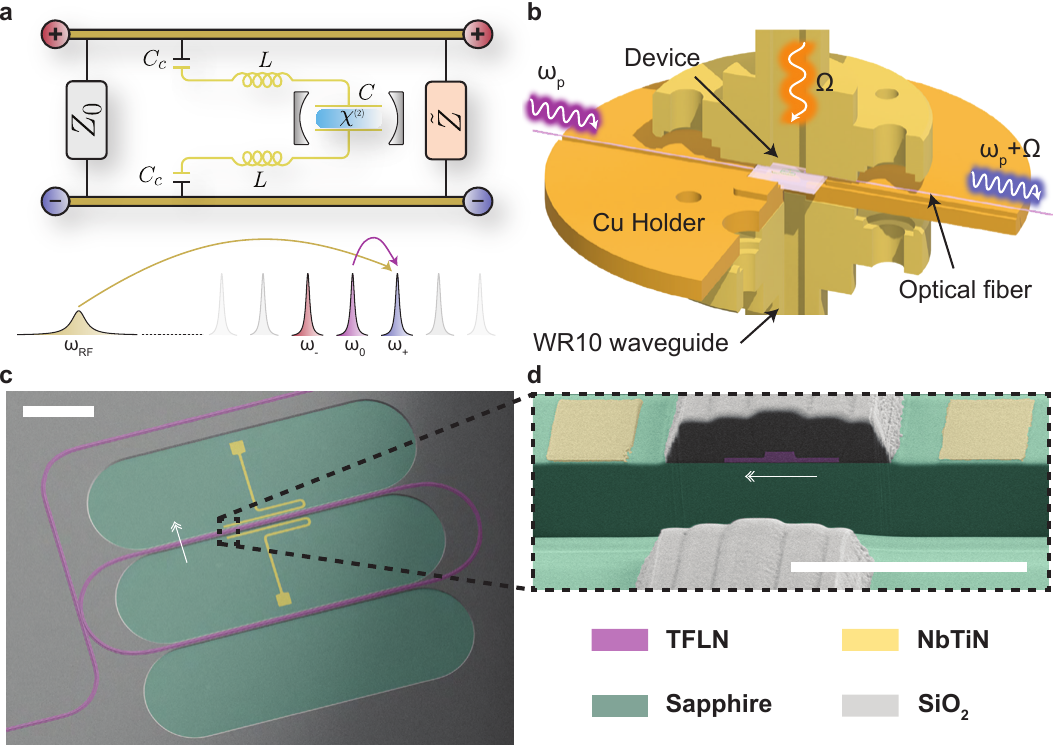}
    \caption{Device operating principle and overview.~\textbf{a} Lumped element model of the transducer. The incoming sub-THz fields are carried by a transmission line with characteristic impedance $Z_0$ and are coupled to the superconducting resonant circuit (colored in gold) by capacitance $C_c$. The outgoing sub-THz waves experience a different characteristic impedance $\tilde Z$, due to the Sa substrate. The intracavity electric field of the sub-THz mode drops across a capacitor of capacitance $C$, which surrounds the LN, depicted as medium with a $\chi^{(2)}$ nonlinearity. The mirrors enveloping the LN crystal indicate that the optical modes are resonant. The cartoon below the circuit model depicts the triply resonant three-wave mixing interaction realized by our device, where $\omega_\text{RF},\ \omega_0,\ \omega_+$ are the sub-THz mode, the optical pump mode, and the up-converted blue-detuned optical sideband mode frequencies. The red-detuned sideband mode, $\omega_-$, is also shown for completeness.~\textbf{b} Rendering of the transducer device and packaging. The highlighted arrows show the directionality of the optical pump (purple, $\omega_{p}$), sub-THz pump (orange, $\Omega$), and up-converted optical sideband (blue, $\omega_{p} + \Omega$), respectively (red sideband, $\omega_p-\Omega$ not shown for brevity). The integrated photonic and sub-THz circuitry lies atop a translucent sapphire substrate and is mounted in the center of the copper holder.~\textbf{c} False-colored scanning electron micrograph of the full device. Scale bar indicates \qty{100}{\um}.~\textbf{d} Zoomed in cross-section of a section of the interaction region. Scale bar indicates \qty{10}{\um}. Double arrowhead indicates the LN crystal-$z$ axis.}
    \label{fig:fig1}
\end{figure*}

Our device utilizes the electro-optic effect, where a sub-THz signal with frequency $\Omega$ modulates the refractive index of a region of \textit{X}-cut thin-film lithium niobate (TFLN), through the electro-optic coefficient, $r_{33}\approx$ \qty{31}{\pm\per\volt}~\cite{weis:1985:lithiumniobate}. In the presence of an optical pump ($\omega_p$), the modulation produces red- and blue-detuned sidebands ($\omega_p\pm\Omega$) via three-wave mixing. By matching the frequency of the sub-THz resonant mode ($\omega_\text{RF}\approx \Omega$) with the free spectral range (FSR) of the optical cavity (\textit{i.e.}, the difference between a telecom optical pump mode $\omega_0$ and the next detuned mode of the optical cavity, $\omega_{\pm} = \omega_0\pm\omega_\text{RF}$), we form a triply-resonant cavity electro-optic system (see~\hyperref[fig:fig1]{Fig. 1a}), which enables efficient pump photon utilization. When the input and generated fields are resonant with their respective cavity modes, the on-chip \textit{quantum} efficiency for photon transduction in the low-cooperativity (cooperativity $C \equiv 4g_0^2n_{c,0}/(\kappa_+\kappa_\text{RF})\ll 1$) regime can be written as,
\begin{align}\label{eq:eq1}
 \eta_\text{OE} &= \left|\frac{\alpha_+^\text{out}}{\beta^\text{in}}\right|^2 \approx 4 C \left(\frac{\kappa_{e,+}}{\kappa_{+}}\right)\left(\frac{\kappa_{e,\text{RF}}}{2\kappa_{\text{RF}}}\right),
\end{align}
where $|\alpha_+^\text{out}|^2 = P_+^\text{out}/(\hbar \omega_+) $ is the (blue) optical photon flux generated on-chip, and $|\beta^\text{in}|^2 = P_{\text{RF}}^\text{in}/(\hbar \omega_\text{RF})$ is the sub-THz photon flux incident to the chip. In these expressions, $n_{c,0}$ is the intracavity photon population of the optical pump mode, $\kappa_{e, j}$ ($\kappa_j$) is the external (total) coupling rate of mode $j$, and $P^{\text{in/out}}_j$ is the on-chip input/output power of mode $j$, where $j\in\{+,-,\mathrm{opt}, \mathrm{RF}\}$. The quantity $g_0$ is the single-photon electro-optic coupling rate, which fundamentally sets the strength of the interaction. Note, due to our double-sided coupling approach for the sub-THz resonator, the coupling efficiency (${\kappa_{e,\text{RF}}}/{2\kappa_{\text{RF}}}$) for the RF resonator can at most be $50\%$.

To implement the triply-resonant system described above, we integrate a superconducting sub-THz resonator made from a thin film of niobium-titanium-nitride (NbTiN) with a photonic racetrack resonator made from TFLN atop a sapphire substrate (Sa). We choose NbTiN because it has a much higher superconducting transition temperature $T_{\rm c}$, and shorter quasiparticle lifetime than aluminum, making it suitable for sub-THz operation. Note, because we use a racetrack resonator, we have a comb of modes with a frequency spacing set by the length of the racetrack, and so there is also a red-detuned optical mode. Although its presence is not detrimental in this work, it could increase the rate of parasitic processes during quantum operation (see Supplementary Information) and should be appropriately addressed.  Methods developed to address this issue in microwave frequency transducers, such as using photonic molecules or mode crossings in multimode resonators, may be used~\cite{holzgrafe:2020:cavityelectrooptics, mckenna:2020:cryogenicmicrowavetooptical, sahu:2022:quantumenabledoperation}.

We refer to the region where the TFLN racetrack resonator is surrounded by superconducting electrodes as the interaction region. The superconducting electrodes form a coplanar stripline resonator where the fundamental mode frequency, $\omega_\text{RF}$ is set by the total electrode length (see Methods). A key aspect of our sub-THz resonator design is that it maintains a unipolar electric field along the interaction region (to avoid cancellations in modulation), without crossing over the integrated photonic components~\cite{shen:2024:photoniclink} (to avoid optical losses due to TFLN proximity to metal and RF losses due to electrode proximity to oxide). A false-colored SEM of the device and an etched cross section of the interaction region are presented in \hyperref[fig:fig1]{Fig.~1c-d}.

We design a custom copper package to simultaneously address the device with both sub-THz and optical fields. The packaging, shown in \hyperref[fig:fig1]{Fig. 1b}, interfaces with both WR10 rectangular waveguides and SMF-28 optical fibers. The optical fibers are aligned and glued to the on-chip grating couplers, maintaining their alignment from \qtyrange{298}{4}{\K}~\cite{mckenna:2019:cryogenicpackaging, wasserman:2022:cryogenichermetically}. We mount WR10 rectangular waveguides to the copper package so that the guided sub-THz field is normally incident to the device. The on-chip superconducting electrodes pick up the incident field, like an antenna, and subsequently modulate the electric field in the interaction region.

%%%%%%%%%%%%%%%%%%%%%%%%%%%%%%%%%%%%%%%%
%%%%%%%%%%%%%%%%%%%%%%%%%%%%%%%%%%%%%%%%

%%========= Fig2 ===============%%
\begin{figure}[]
    \centering
    \includegraphics[width=\linewidth]{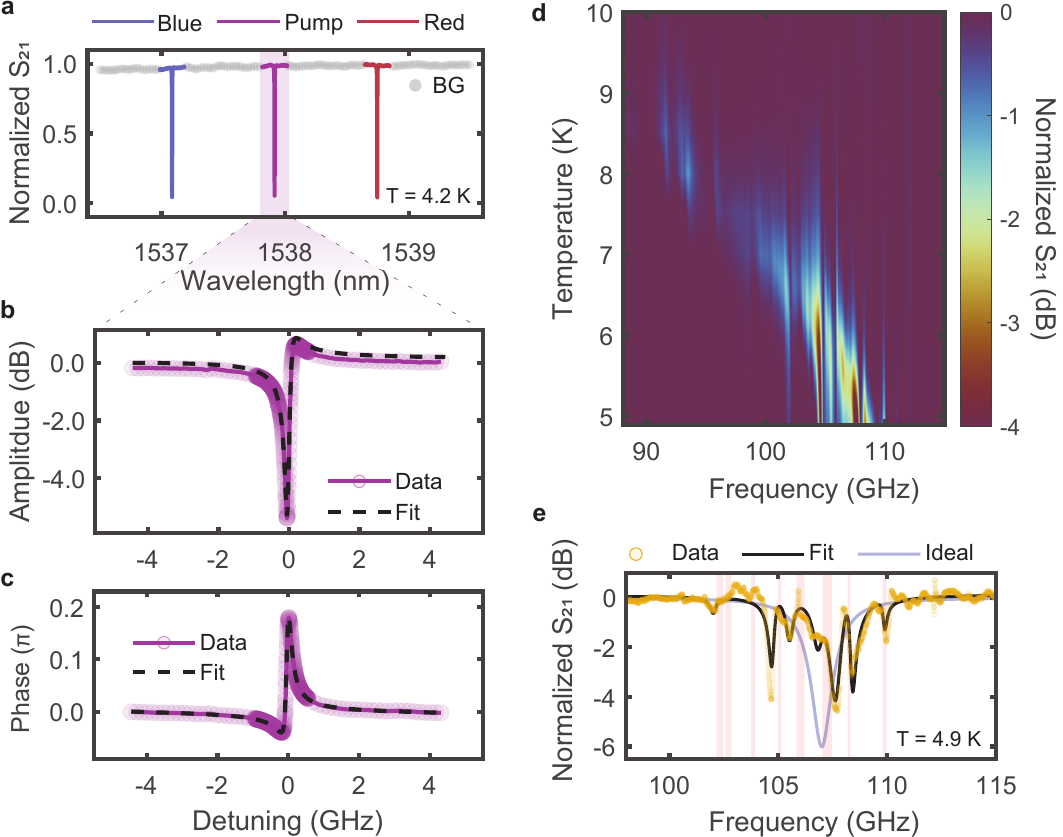}
    \caption{Cryogenic optical and sub-THz resonator characterization.~\textbf{a} Normalized transmission of the three optical modes relevant in our experiments. Purple indicates the pump mode, $\omega_0$, blue (red) indicates the blue- (red-) detuned sideband mode, $\omega_+$ ($\omega_-$). While the distance from $\omega_0$ to the red or blue modes is slightly different due to dispersion, both are within a few MHz of \qty{105.25}{\GHz}.~\textbf{b} Self-heterodyne amplitude response of the optical pump mode, measured on a VNA. We only show the pump mode here for brevity, but we perform these measurements for the two sideband modes as well. \textbf{c} Self-heterodyne phase response of the optical pump mode, measured on a VNA. We fit the phase response to infer the optical mode resonant frequencies $\omega_j$, external coupling rates $\kappa_{e,j}$, and the total loss rates $\kappa_{j}$. These parameters are then used to plot the fits depicted in (b-c).~\textbf{d} Sub-THz transmission spectrum as a function of cryostat platform temperature. We normalize the spectrum to non-superconducting data where $T>T_\mathrm{c}$. We observe the sub-THz mode red-shifting and broadening as temperature increases. Note the presence of avoided crossings from the sub-THz mode's hybridizing with many parasitic Sa substrate modes.~\textbf{e} Single temperature slice from subfigure (d), at $T\approx$~\qty{4.9}{\K}, shown in gold. The solid black line indicates the fit to a multi-parameter model that includes the substrate modes (see Methods, Supplementary Information). We indicate the resonant frequencies of the substrate modes by transparent red vertical lines, where the width indicates the substrate mode linewidth. The light blue, transparent line shows the line-shape of the sub-THz mode that hybridizes with these substrate modes.}
    \label{fig:fig2}
\end{figure}
%%==================================%%

\subsection*{Optical \& sub-terahertz cavity spectroscopy}
We mount the fully packaged device inside a cryostat (Montana Instruments) and cool to a base temperature of roughly $\qty{4.9}{\K}$. Before operating the transducer, we separately characterize the properties for each optical and sub-THz cavity mode. Beginning with the optics, we record an optical spectrum, depicted in~\hyperref[fig:fig2]{Figure 2a}. By calibrating the wavelength axis of this spectrum using a fiber Mach-Zehnder interferometer (MZI), we infer a mode spacing of $|\omega_{\pm} - \omega_0| \approx 2\pi\cdot\qty{105.25}{\GHz}$. Furthermore, we observe a roughly $3$~dB decrease in transmission through the fibers and gratings after the base temperature is reached, which we attribute to temperature-induced changes in the device refractive indices, leading to shifted peak frequency responses of the grating couplers.

Using an off-chip electro-optic modulator (EOM) and vector network analyzer (VNA), we perform heterodyne measurements to further characterize each optical mode. These measurements let us directly infer the external and total coupling rates, $\kappa_{e,j}$ and $\kappa_j$, of each mode. In this measurement, we lock the pump laser blue-detuned of each mode and then, by driving the EOM with the VNA source port, sweep a sideband across the optical resonance. The chip's output is collected on a high-speed photodiode, and the beat tone between the locked pump and the swept sideband is recorded on the receiving VNA port. We fit an input-output model to the phase response of this measurement (see Methods and Supplementary Information). We observe an amplitude and phase response, which are shown in~\hyperref[fig:fig2]{Fig.~2b-c}, respectively. We summarize the inferred optical mode parameters from these measurements in~\hyperref[tab:extended_mode_params]{Extended Data Table 1}.

We characterize the sub-THz resonator using a VNA with frequency extension modules to measure the response from \qtyrange{75}{115}{\GHz}. Because the sub-THz mode hybridizes with many parasitic modes, it is difficult to distinguish the sub-THz mode from the background. To identify superconducting features, we perform a temperature sweep by changing the platform temperature of the cryostat and measuring the sub-THz spectrum at each temperature. We observe a broad feature that tunes with temperature, which we identify as the sub-THz mode for further processing (see~\hyperref[fig:fig2]{Fig. 2d}). All sub-THz data shown in~\hyperref[fig:fig2]{Figure 2} are normalized to a spectrum taken at $T>T_\mathrm{c},\ T_\mathrm{c}\approx \qty{10.5}{\K}$.

We observe the superconducting sub-THz mode red-shift and broaden, confirming the predicted response of superconducting resonators to thermal excitation. The parasitic modes are largely unaffected by the \qty{5}{\K} temperature change, indicating that they are not superconducting. From this observation, we infer that they are substrate modes. Consequently, we develop a multiparameter input-output model of the hybridization, which we use to fit the sub-THz data (see Methods and Supplementary Information). Modeling the hybridization is important because it impacts the estimate of the intracavity RF photon number, which impacts our estimate for $g_0$, the electro-optic coupling rate (see Equation 2, Methods, and Supplementary Information). An example RF spectrum, along with a fit to our multiparameter model, is shown in \hyperref[fig:fig2]{Fig.~2e}. As a result of including the substrate modes in our model, we obtain self-consistency between the two different methods we use to estimate $g_0$ (discussed in the next section). We summarize the sub-THz mode parameters, the substrate mode parameters, and the relevant coupling rates, in~\hyperref[tab:extended_mode_params]{Extended Data Table 1}.

\subsection*{Transducer operation}
\begin{figure*}[h!]
\centering
\includegraphics[width=\linewidth]{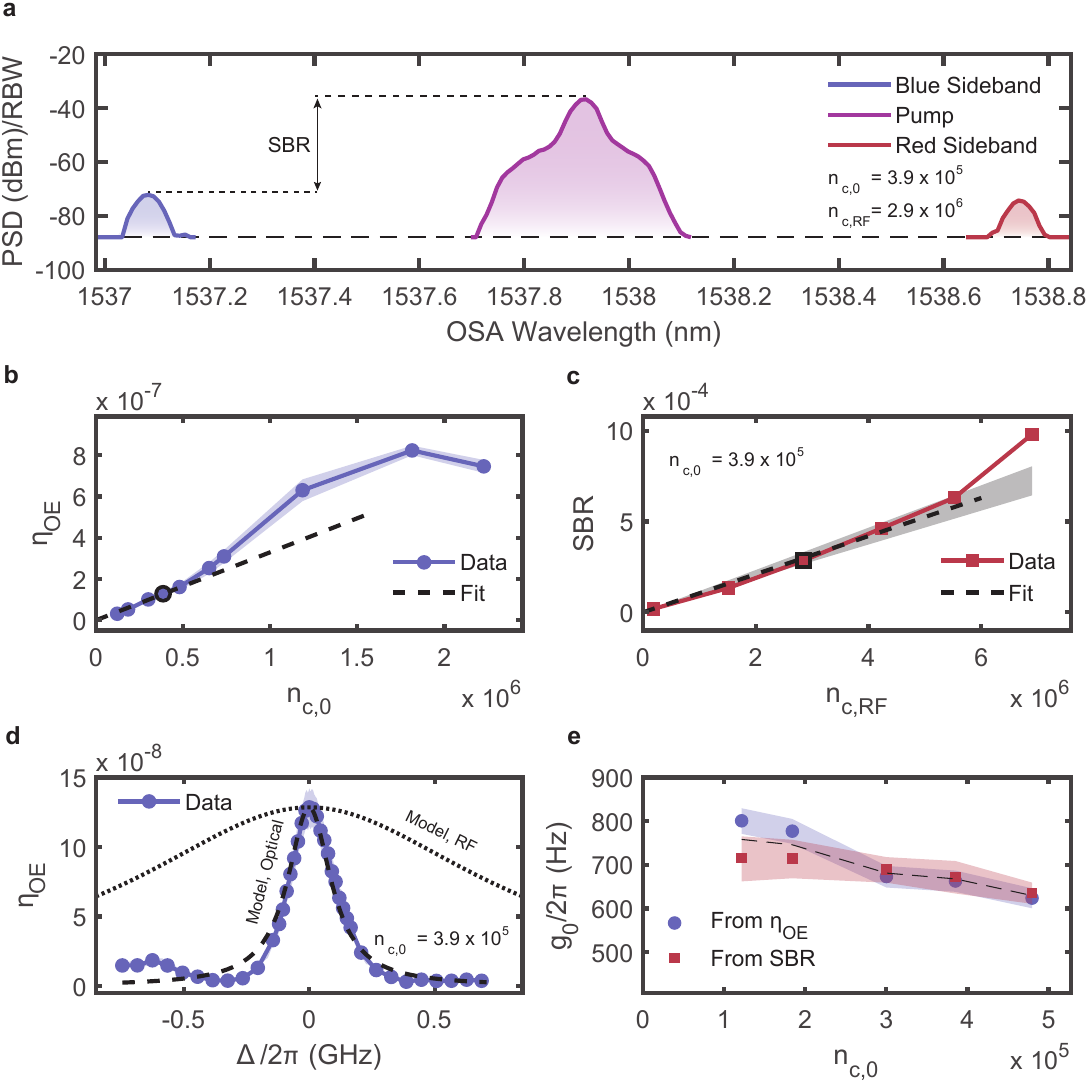}
\caption{Transducer characterization at $T\approx\qty{4.9}{\K}$.~\textbf{a} Raw power spectral density data as measured from the OSA for modulation frequency $\Omega = 2\pi\cdot\qty{105.285}{\GHz}$. In this data the optical and RF power corresponds to $n_{c,0} = 3.9\times10^5$ and $n_{c,\text{RF}}=2.9\times10^{6}$ resulting in a transduction efficiency, $\eta_\text{OE}\approx 1.3\times10^{-7}$ and a sideband ratio, $\Upsilon_{+,0} = 2.9\times10^{-4}$; the highlighted blue circle and red square in subfigures (b) and (c), respectively. The resolution bandwidth of the OSA is \qty{0.05}{\nm}.~\textbf{b} The transduction efficiency as a function of intracavity optical pump photon number, where the shading indicates 95\% confidence interval. The dashed line depicts a linear fit to the low power data, from which we can infer $g_0 \approx 2\pi \cdot \qty{0.7}{\kHz}$.~\textbf{c} Side-band-ratio (SBR) versus intracavity sub-THz photon number. This is the ratio of the transmitted blue sideband power to the transmitted optical pump power. By fitting a line to the low-RF-power data (dashed line), we can extract a separate measure of $g_0$ which agrees well with the results in subfigure (b). The grey shading indicates a 95\% confidence interval to the linear fit.~\textbf{d} Transduction efficiency as a function of modulation frequency ($\Omega$) for the highlighted blue circle in subfigure (b). The long-dashed Lorentzian is not determined by a fitting procedure, but instead by centering a Lorentzian function at the peak modulation frequency and setting its width to the independently measured blue mode optical linewidth. A similar Lorentzian model with the measured sub-THz linewidth is plotted as the short-dashed line. The shading indicates a 95\% confidence interval.~\textbf{e} Comparison of $g_0$ inferred from two analysis methods. Blue circles indicate values inferred from the loss-calibrated transduction efficiency (i.e. subfigure (b)). Red  squares indicate values inferred from the self-calibrated SBR (i.e. subfigure (c)). Transparent lines show the 95\% confidence interval at each optical power. The dashed line depicts the mean of the two curves.} 
\label{fig:fig3}
\end{figure*}

In order to measure transduction, we first lock our laser close to the pump-mode resonance frequency, $\omega_0$ (see Methods). We maintain a lock to the pump mode via PID loop feedback on the laser piezo voltage implemented with a Red Pitaya. Slight fluctuations in the lock-point voltage can vary the detuning of the laser from the resonance, so we fit this detuning as a parameter in post-processing (see Methods and Supplementary Information). The laser wavelength is locked blue of the cavity resonance to increase stability against thermo-optic and photorefractive mode drift (see Supplementary Information). Sub-THz modulation at frequency $\omega_{\text{RF}} \approx |\omega_{\pm}-\omega_0|$ generates optical sidebands that are visible and measured directly on an optical spectrum analyzer (OSA). An example OSA trace is shown in~\hyperref[fig:fig3]{Fig.~3a}. The three peaks correspond to the optical modes $\omega_0$ and $\omega_{\pm}$ shown in ~\hyperref[fig:fig2]{Fig.~2a}. Note the difference between the powers in the red and blue sidebands. This asymmetry arises from different internal and external coupling rates in the two modes (see Supplementary Information). 

We characterize the device transduction efficiency $\eta_\text{OE}$ and single-photon coupling rate $g_0$ by separately varying both the optical pump power and RF modulation power. Each optical pump power yields a different intracavity photon number in the mode $\omega_0$. By also varying the RF modulation power, we obtain multiple measurements on the OSA, from which we determine the on-chip photon flux in both the blue sideband and the RF cavity. We thereby obtain a linear plot of the on-chip blue photon flux, $\dot{n}_+$, versus the incident RF photon flux, $\dot{n}_{\mathrm{RF}}$. The slope of this line, $\eta_\mathrm{OE} = \dot{n}_\mathrm{+}/\dot{n}_\mathrm{RF}$, is the electrical-to-optical number conversion efficiency, and constitutes a single data point in the plot in \hyperref[fig:fig3]{Fig.~3b}. More details of this procedure are available in Methods and Supplementary Information.

\begin{figure}[h!]
\centering
\includegraphics[width=\linewidth]{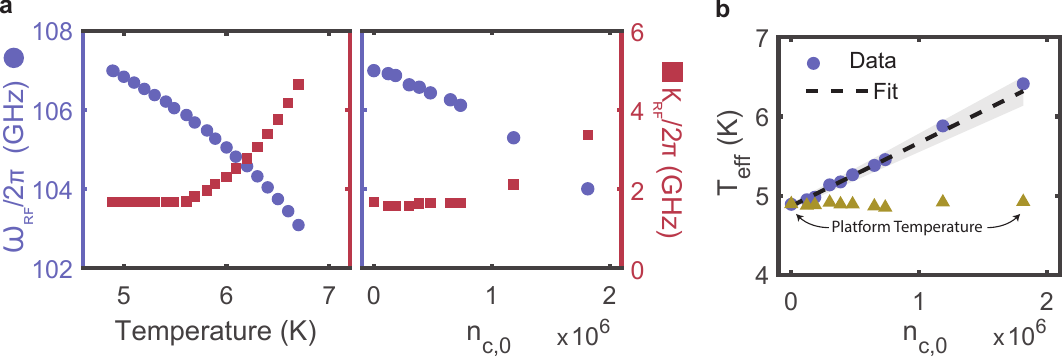}
\label{fig:fig4}
\caption{Sub-THz frequency and linewidth as a function of local heating.~\textbf{a} Sub-THz resonance frequency and linewidth versus cryostat platform temperature (left plot) and versus intracavity optical pump photon number (right plot). Blue circles indicate frequency and red squares indicate linewidth. We observe similar shifts in frequency and linewidth between the two subfigures. We ascribe the shifts resulting from increased intracavity photon number to local heating of the sub-THz superconducting resonator.~\textbf{b} Effective temperature of the sub-THz resonator versus intracavity optical pump photon number. The effective temperature is determined by interpolating the temperature data, for a given detuning and optical photon number. The black-dashed line is a linear model, fit to the data, with the grey shading showing the 95\% confidence interval. We determine the slope and corresponding 95\% confidence interval to be \qty{0.80}{\micro\K\per photon} (\qty{0.72}{}, \qty{0.88}{})\space\qty{}{\micro\K\per photon}. Beige triangles indicate the cryostat platform temperature. The constancy of the platform temperature indicates that the effective temperature increase is predominantly a local effect.}
\end{figure}

In the low-cooperativity limit, the on-chip transduction efficiency should increase linearly with the intracavity optical photon number $n_\mathrm{c,0}$, as indicated by Eq.~\ref{eq:eq1}. We observe the expected linear trend at low intracavity pump photon numbers. However, as the intracavity photon number increases, we see a deviation in the efficiency curve (\hyperref[fig:fig3]{Fig.~3b}). We attribute this effect to local heating of the sub-THz resonator by absorption of pump photons that changes the tuning condition (see~\hyperref[fig:fig4]{Fig. 4}, Methods and Supplementary Information). The effect can be replicated in our model by including the optical-power-dependent detuning and linewidth of the RF mode (see~\hyperref[fig:fig4]{Fig. 4}) into~\hyperref[eq:eq1]{Equation 1}.

Inferring the single-photon coupling rate $g_0$ from the measured transduction efficiency hinges on an accurate estimate of the on-chip photon numbers in both the optical and sub-THz resonators. Achieving this requires careful calibration of the losses in each measurement chain, along with precise power monitoring at cryogenic temperatures. Even minor misalignments, temperature drifts, or repeated unplug/replug cycles can lead to shifts in the optical throughput or in the power delivered to the chip, which in turn directly affect the inferred values of $g_0$. We therefore perform cryogenic calibrations of both the optical and RF paths to minimize these sources of error, as described in the Supplementary Information.

In addition to measuring the transduction efficiency, we perform an analogous measurement of a sideband power \textit{ratio} (SBR) that significantly reduces the sensitivity to optical power calibration. The SBR is defined as $\Upsilon_{+,0}\equiv P_+/P_0$, where $P_j$ refers to the pump power ($P_0$) and the blue sideband power ($P_+$), as measured by the OSA. On resonance, the SBR can be written as
\begin{equation}
    \label{eq:eq2}
    \Upsilon_{+,0} \approx \frac{4g_0^2 }{\kappa_+^2}\cdot \frac{4\kappa_{e,0}\kappa_{e,+}}{(\kappa_0-2\kappa_{e,0})^2} \cdot n_{c,\text{RF}},
\end{equation}
where $n_{c,\text{RF}} = 2\kappa_{e,\text{RF}}/\kappa_{\text{RF}}^2\cdot \frac{P_\text{RF}}{\hbar\omega_{\text{RF}}}$. The SBR is directly proportional to $g_0^2$ as in $\eta_\text{OE}$, but instead of scaling with the number of intracavity optical pump photons $n_{c,0}$, the SBR scales with the number of intracavity RF photons $n_{c,\text{RF}}$ (see Methods and Supplementary Information). Any optical losses in the detection chain normalize out of this expression, which therefore reduces the influence of optical path calibration errors on our estimate of $g_0$. 

The inferred values of $g_0$ from both methods for low pump power (corresponding to the linear region in~\hyperref[fig:fig3]{Fig. 3b}) are shown in \hyperref[fig:fig3]{Fig.~3e}. In principle, the inferred $g_0$ should be constant for each pump power, for both methods. Experimental factors such as polarization drift, inaccurate path calibration for the optics and RF, and uncertainties in modal rates ($\kappa$'s) and detunings can cause discrepancies. In our case, we find fairly good agreement across all pump powers, indicating that our measurements are consistent within the statistical uncertainty. If we average all the estimates of $g_0$ from the $\eta_\text{OE}$ data, we obtain $g_0^\eta = 2\pi\cdot 707 \text{ } (681, 733) \text{ Hz}$, where the parentheses indicate the 95\% confidence interval. Likewise, for the estimates using the SBR we obtain $g_0^{\Upsilon} = 2\pi\cdot 685 \text{ } (645, 722) \text{ Hz}$.

To characterize the bandwidth of the transducer, we sweep the modulation frequency ($\Omega$) for each optical and RF power. Representative data from bandwidth measurements are shown in~\hyperref[fig:fig3]{Fig.~3d}. We observe a maximum conversion efficiency at $\Omega = 2\pi\times\qty{105.285}{\GHz}$ ($\Delta=0$), in agreement with the measurement of the MZI-calibrated free spectral range discussed above, $|\omega_+-\omega_0| = 2\pi\times\qty{105.25}{\GHz}$. Together with the data in~\hyperref[fig:fig3]{Figure 3d}, we plot two sets of dashed lines. The long-dashed line depicts a Lorentzian model, with a peak value chosen to match the data and a linewidth chosen to correspond to the measured optical mode linewidth of the blue sideband mode at $\omega_+$. In contrast to previous work on optomechanical and electro-optic transducers, our conversion bandwidth is limited by the \textit{optical cavity linewidth}, which for us is the smallest rate in our system. This puts us in the so-called reversed dissipation limit of cavity electro-optics~\cite{nunnenkamp:2014:quantumlimitedamplification}. We also show a short-dashed line, depicting a Lorentzian model of the sub-THz mode, again with the peak chosen to match the data and with a linewidth corresponding to the inferred superconducting sub-THz mode linewidth as reference. Note that there is a small side peak in the conversion efficiency at $\Delta \approx -2\pi\times 0.6~\text{GHz}$, which we attribute to hybridization with a substrate mode (see Methods).

%%%%% HEATING DISCUSSIONS %%%%%%%%%%%
%%========= Fig 4 ===============%%
%%===============================%%

\subsection*{Locally heating the sub-THz resonator}
As mentioned previously, we attribute the nonlinear behavior in efficiency to heating of the superconducting sub-THz resonator (see~\hyperref[fig:fig3]{Fig.~3b}). As the number of optical pump photons increases, more photons are scattered into the environment and eventually are absorbed by the sub-THz electrodes, thereby shifting and broadening the superconducting resonance. To quantify this effect, we define an effective temperature by matching the sub-THz shifts due to the cryostat temperature and pump intracavity photon number, shown in~\hyperref[fig:fig4]{Figure 4c}.

Fitting the normalized temperature sweep data from \hyperref[fig:fig2]{Fig.~2d} to our multiparameter model, we extract the sub-THz mode frequency and linewidth at each temperature and plot the results in \hyperref[fig:fig4]{Fig.~4a}. We then incorporate the substrate mode parameters into a similar multi-parameter fit of the RF spectrum recorded at each input optical pump power (normalized to the same $T>T_\mathrm{c}$ background). This is plotted in \hyperref[fig:fig4]{Fig.~4b}.

By comparing the impact of the cryostat platform temperature and the optical pump power on the frequency and linewidth of the sub-THz mode (~\hyperref[fig:fig3]{Figure~3b}\hyperref[fig:fig3]{, d}), we construct an ``effective temperature''. We interpolate the data to identify the effective temperature corresponding to each pump photon number, shown in~\hyperref[fig:fig4]{Fig.~4c}. Because the measured platform temperature remains nearly constant for each intracavity pump photon number, we conclude that the effective temperature increase is due to local heating caused by the continuous flux of pump photons. 

%%==================================%%
%%         DISCUSSION               %%
%%==================================%%

\section*{Discussion}
\label{discussion}
In summary, we have demonstrated an integrated sub-THz cavity electro-optic transducer on thin-film lithium niobate. Our maximum observed efficiency is approximately $0.82 \times 10^{-6}$ with a $3$-dB bandwidth of $\kappa_+ \approx 2\pi\cdot210$~MHz, closely following the linewidth of the optical mode. Furthermore, we infer a single-photon coupling rate of $g_0 \approx 2\pi\cdot0.7$~kHz. To understand and improve the performance of this device, we need to consider some key factors: fabrication, substrate mode mitigation, dispersion engineering, and operation. Note that being able to tune the photonic and/or the sub-THz resonances greatly improves ease of operation and fabrication yield which can be addressed in tandem with the discussion below. 

%%%% FABRICATION CONSIDERATIONS / Q ENHANCEMENTS %%%%
Our current fabrication process for the sub-THz resonator requires us to protect the TFLN racetrack by depositing oxide cladding atop the chip. However, we find that this oxide increases the loss of the sub-THz mode. In addition to lowering the quality factor of the sub-THz mode, the oxide cladding forces us to position the NbTiN electrodes farther away from the LN than necessary. This increased distance reduces the single-photon coupling rate, $g_0$, and thus the performance of the device. Therefore, we expect that eliminating the oxide cladding from our process would immediately improve the device. Some options for doing this would be to utilize lift-off instead of etching to define the NbTiN resonator or to use flip-chip methods to place the superconducting resonator closer to the optics. 

%%%%% SUBSTRATE MODES %%%%%
The sub-THz substrate modes supported by the Sa substrate reduce the overall performance of our device. These substrate modes hybridize with the fundamental superconducting mode, thereby changing the electric field distribution. Consequently, there are two main effects: the field overlap between the RF and optical modes could reduce, reducing $g_0$; and the intracavity sub-THz photon number is modified by the introduction of additional detuning and broadening of the sub-THz mode (see Methods). Additionally, as seen in~\hyperref[fig:fig2]{Figure~2d}, these modes complicate the fitting of the sub-THz spectrum. We envision two possible solutions to overcome the effects of substrate modes. First, by reducing the volume of the sapphire substrate (\textit{i.e.}, by changing the chip dimensions from \qty{4.6}{\mm} $\times$ \qty{2.37}{\mm} $\times$ \qty{0.5}{\mm} to \qty{2.7}{\mm} $\times$ \qty{1.3}{\mm} $\times$ \qty{0.1}{\mm}) we can reduce the number of substrate modes from 29 to 1 in the \qty{95}{\GHz} -- \qty{110}{\GHz} frequency range (see~\hyperref[fig:ef3]{Extended Figure 3}). Secondly, through careful engineering, we could redesign the packaging to guarantee that there is minimal coupling to substrate modes. Modifying the low-loss coupling structure in~\cite{anferov:2024:lowlossmillimeterwave, anferov:2024:millimeterwavesuperconducting} to accommodate TFLN photonics provides a platform that natively integrates with current millimeter-wave/sub-THz superconducting qubit packaging.

%%%%% OPERATION \& LOCAL HEATING %%%%%
Improving the device's thermal operation is critical for quantum applications. All of the experiments discussed in this paper are carried out with a CW optical pump, which effectively bathes the superconducting electrodes in a constant stream of photons. Absorption of these photons is physically equivalent to increasing the temperature of the electrodes, as shown in~\hyperref[fig:fig4]{Fig. 4c}. The photons are locally heating the superconducting electrodes, increasing its quasiparticle density. While the frequency of the superconducting resonance decreases, the linewidth of the resonance increases, with both effects harming the device performance (see~\hyperref[fig:fig3]{Fig. 3b}). A common technique used for microwave-to-optical transducers is to pulse the pump. By pulsing the pump, time is provided for the quasiparticle density to relax to its thermal equilibrium value. This enables ideal superconducting conditions for each transduction event (mediated by the pump-pulse). 

Furthermore, in this work, we operate the device at roughly \qty{5}{\K}. At this temperature, the average thermal photon occupation is approximately 0.6 per mode at \qty{105}{\GHz}. To perform any quantum experiment, we would need to operate at a lower temperature. For the same frequency, operating at \qty{1}{\K}, the thermal occupation is less than 0.01 photons per mode, which is sufficiently low~\cite{anferov:2024:millimeterwavesuperconducting}. Operating at a lower temperature comes with the additional benefit that the superconducting RF resonator will have an increased quality factor.

%%%%%%%%%%% CONCLUSION %%%%%%%%%%%%%%%%%%%%%
Overall, we have implemented a sub-THz cavity electro-optic system and demonstrated coherent sub-THz-to-optical transduction. Our work shows that integrated devices operating at sub-THz frequencies are not only feasible, but potentially advantageous for future quantum applications by enabling higher-temperature operation. With a clear path forward to improve our device, we envision a new regime of quantum superconducting technologies operating at sub-THz frequencies. Our work ultimately highlights the challenges inherent to designing integrated sub-THz photonic systems and provides a path toward future advancements in sub-THz quantum hardware. 
%TC:ignore

%%==================================%%
%%           REFERENCES             %%
%%==================================%%

%%==================================%%
%%             METHODS              %%
%%==================================%%
\clearpage
\section*{Methods}
\label{sec:methods}
\renewcommand{\theequation}{M\arabic{equation}}

%%%%% SUBSECTION: EXPERIMENTAL SETUP %%%%%
\subsection*{Experimental setup}
\label{sec:fullpath}

The complete experimental setup, depicted in~\hyperref[fig:ef1]{Extended~Figure~1}, can be separated into two optical and two electronic signal paths. These paths comprise: the primary optical path for optical sideband measurements; a secondary path for self-heterodyne measurements (\hyperref[fig:fig2]{Figure~2b, c}); the primary sub-THz RF path; and the electronic signal detection path, used to lock the optical pump to the mode $\omega_0$. Please see supplementary information for detailed descriptions of these different paths. \hyperref[fig:ef4]{Extended~Figure~4c} presents an optical image of the inside of the cryostat, showing the WR10 waveguides and part of the sub-THz waveguide network.

%%%%% SUBSECTION: PUMP DETUNING / LOCKING %%%%%
\subsection*{Pump detuning calibration and locking}
\label{sec:lock}
We lock our laser to the optical cavity using a Red Pitaya in combination with open source software, PyRPL. We control our laser's (Santec TSL-710) wavelength via piezo voltage control, which is set by either the Red Pitaya output or from our DAQ's analog output directly; selected using a mechanical BNC relay. Using PyRPL, we specify a voltage corresponding to transmission slightly detuned from resonance of the optical cavity pump mode and monitor the output transmission as a voltage on our APD. This voltage is routed to the Red Pitaya, and the PID loop varies the piezo voltage sent to the Santec to maintain a constant transmission voltage. We tune into the PID lock point from the blue side of the mode, which we observe is a more stable operating point than the red-detuned side. Upon initially locking to the mode, the optical cavity shifts dramatically as a result of the thermo-optic and photorefractive effects in TFLN~\cite{xu:2021:mitigatingphotorefractive1}. We find that after pumping the cavity continuously for some time, the rate of cavity drift drops significantly, which we attribute to charge carriers saturating trap sites until we reach a steady state (see Supplementary Information for more details).

The exact detuning of the locked pump from the optical resonance ($\Delta_0$) is critical to our analysis, but direct measurement is difficult. This value is obtained through a single parameter fit of the transmission $T = |1 - \frac{\kappa_{e,0}}{i\Delta_0 + \kappa_0/2}|^2 = P_\text{OSA}/P_\text{PM} \cdot \eta_0$, where $T$ is the transmission through the optical pump cavity, $P_\text{OSA}$ is the measured optical pump power on the optical spectrum analyzer at the pump wavelength, $P_\text{PM}$ is the optical power reading from a power meter right before the laser is sent into the cryostat, and $\eta_0$ is a measured factor, which encapsulates measurements of path losses, beam-splitter ratios, grating coupler efficiency, and differences in calibration between the OSA and PM. We infer $\kappa_{e,0}$ and $\kappa_0$ from self-heterodyne measurements, as described in the main text (more details provided in Supplementary Information).

%%%%% SUBSECTION: DEVICE FAB / PACKAGING %%%%%
\subsection*{Device fabrication \& packaging}
\label{sec:fab}
Our fabrication closely follows that of McKenna \& Witmer et al.~\cite{mckenna:2020:cryogenicmicrowavetooptical}. We start with a \qtyproduct{12 x 16}{\mm} piece of material with nominally \qty{500}{\nm} of MgO-doped $X$-cut LN atop a \qty{500}{\um}-thick $C$-cut Sa. The optical waveguides, the racetrack resonator, and the grating couplers are defined using hydrogen silsesquioxane (HSQ) resist and electron beam lithography (JEOL, JBX-6300 FS, 100 keV). Using an Intlvac Ion Beam Mill (argon ions), we etch roughly \qty{300}{\nm} of TFLN to transfer the mask into the TFLN. 

Next, we etch excess TFLN around the optical components. We define a second mask using SPR3612 resist and photolithography (Heidelberg MLA 150, \qty{375}{\nm}) and etch the remaining roughly \qty{200}{\nm} of TFLN slab around the optical devices via the Intlvac ion mill. At this stage, we clean the sample using various acids, followed by annealing at \qty{500}{\degree\C} in atmosphere.

We next clad the chip with approximately \qty{2}{\um} of silicon dioxide (SiO$_2$) using low-temperature high-density chemical vapor deposition (PlasmaTherm Versaline HDP-CVD), to protect the optical components from NbTiN deposition in a later step. Cladding takes two steps of roughly \qty{1}{\um} depositions, separated by a \qty{530}{\degree\C} anneal in atmosphere~\cite{shams-ansari:2022:reducedmaterial}. As discussed in the main text, SiO$_2$ contributes to loss in the sub-THz resonance, so we etch windows into the oxide where we intend to place the sub-THz resonator. Using SPR220-3 resist, we pattern windows via another round of photolithography and etch the SiO$_2$ with fluorine chemistry in an inductively coupled plasma reactive ion etcher (PlasmaTherm ICP RIE). With the Sa exposed in these regions, we deposit NbTiN via DC magnetron co-sputtering of niobium (Nb) and titanium (Ti) in a mixed nitrogen-argon environment (Kurt J. Lesker PVD Pro-line).

In a third round of photolithography, we pattern the sub-THz resonators with SPR3612 resist. Lastly, we etch the NbTiN with a mixture of SF$_6$/Ar (PlasmaTherm ICP RIE). This device is pictured in \hyperref[fig:ef4]{Extended~Figure~4a, b}.

%%%%% SUBSECTION: SUB-THZ DEVICE DESIGN %%%%%
\subsection*{Sub-terahertz superconducting resonator design}
\label{sec:subthzdesign}
We design the sub-THz resonator such that it maintains a predominantly single-polarity voltage drop across the TFLN. Due to RF loss in the SiO$_2$ cladding, we cannot cross the oxide with the resonator and therefore carefully design a half-wave-like resonator. We modify a half-wave transmission coplanar stripline resonator so that the polarity of the electric field does not change sign between the electrodes in the TFLN region (see Supplementary Information for additional details). We present the current density distribution and the electric field distribution of the mode in \hyperref[fig:ef2]{Extended~Figure~2a-c}.

The dimensions of the sub-THz device dictate its resonant frequency given a fixed kinetic inductance. Predominantly, the resonator length ($L$) determines its fundamental frequency, whereas the length of the coupling section ($L_c$) and size of the capacitive pads control the external coupling rate of the resonator to the WR10 waveguide (labeled in \hyperref[fig:ef2]{Extended~Figure~2a}). We model the fundamental frequency as $\omega_\text{RF}(L) = v_p/L + \delta\omega$, where $v_p$ is the phase velocity of the coplanar stripline section, and $\delta\omega$ describes a shift in resonance frequency due to the capacitive coupling region (the pads and coupling section).

To incorporate the effects of kinetic inductance in the design, we estimate the sheet inductance using the relationship $L_\text{s} \approx \hbar/(\pi \Delta_0) \cdot R_\text{s}$. In this expression, $L_\text{s}$ is the kinetic sheet inductance, $\Delta_0 \approx 1.764 T_\mathrm{c}$ describes the superconducting gap at T=\qty{0}{\K} of the material, and $R_\text{s}$ describes the film's normal-state sheet resistance. During fabrication of the main device, we include a witness sample so that we can measure the film sheet resistance and thickness, and infer the resistivity. For the device presented in this paper, we measure $\rho_s \approx \qty{250}{\micro\Omega\cdot\cm}$ for a \qty{50}{\nm} film on the witness chip. From here, we can estimate $L_s\approx \qty{6.6}{\pico\henry}/\square$ assuming $T_\mathrm{c}\approx \qty{10.5}{\K}$. With this estimate of the sheet inductance, we design superconducting test devices (using SONNET Software) and measure them before continuing fabrication on the main device. We summarize the measurements of the NbTiN-witness devices against our simulation of the expected resonance frequencies in~\hyperref[fig:ef2]{Extended Fig. 2d}. The relationship between the measured resonance and simulated resonance frequencies can be shown to be linear,
\begin{equation}
    \omega_\text{RF}^\text{actual} = r \omega_\text{RF}^\text{sim} + \epsilon.
\end{equation}
Using this model, we can design and pattern the NbTiN resonators on the transducer device. We sweep the design parameters of the sub-THz resonator across the fabricated devices to increase robustness against film and optical device variations. Based on the measured optical and sub-THz characteristics of the various fabricated devices, we select one for detailed study in this manuscript. Note that after incorporating real device dimensions (as measured from scanning electron microscopy), we use SONNET Software to infer the sheet inductance to be $L_s^\text{meas}\approx \qty{7.6}{\pico\henry}/\square$, by matching the simulated resonant frequency to the measured one; this sheet inductance value is within 15\% of our estimate based on the BCS theory.

\subsection*{Substrate mode modeling \& analysis}
\label{substratemodel}
The approximately \qty{500}{\um} thick substrate, which is the standard thickness for Lithium Niobate-on-Sapphire (LiSa) wafers, acts as a cavity that supports many sub-THz modes. To quantify these substrate modes, we perform eigenmode simulations in COMSOL and count the number of modes between \qty{95}{\GHz} and \qty{110}{\GHz}, as a function of geometry. In~\hyperref[fig:ef3]{Extended Fig. 3a} we show the geometry of our simulation, where all boundaries are perfect electrical conductors (PEC) except the waveguide input and output ports. As the thickness of the sapphire decreases, the number of supported modes also decreases. Some, but not all, of these substrate modes are visible in the sub-THz resonator spectrum (see \hyperref[fig:fig2]{Figure~2e}). The frequency domain simulation in~\hyperref[fig:ef3]{Extended Fig. 3d} exhibits only a handful of modes, though we predict 29 eigenmodes to be supported. This suggests that only modes with polarization corresponding to the TE10 mode of the WR10 waveguide are driven. 

Separately, in our model, we assume that the substrate modes are not directly accessible or driven via the WR10 waveguide field. Instead, only substrate modes with an appropriately matched polarization and spatial mode profile can couple to the superconducting resonator and are visible in the measured RF spectrum~(see \hyperref[fig:fig2]{Figure~2e}). This coupling, which hybridizes the superconducting RF mode with the substrate modes, leeches RF photons from the superconducting cavity. The result of this hybridization is an effective loss increase and detuning of the superconducting sub-THz mode. The RF spectrum accounting for these hybridized substrate modes is then described by: 
\begin{equation}\label{eq:m4}
    S_{21}^\text{RF} = 1 - \frac{\kappa_{e,\text{RF}}/2}{i(\Delta_\text{RF}+\tilde\delta) + (\kappa_\text{RF}+\tilde\gamma)/2 }
\end{equation}
From Equation M4, we see the effects of the substrate modes acting on the superconducting resonance are additional loss and detuning described by:
\begin{align}
    \tilde\delta &= -\sum_{n=1}^N \frac{|J_n|^2\Delta_n}{\Delta_n^2+\gamma_n^2/4} \\
    \tilde\gamma/2 &=  \sum_{n=1}^N \frac{|J_n|^2\gamma_n/2}{\Delta_n^2 + \gamma_n^2/4},
\end{align}
where $J_n$ is the coupling rate, $\Delta_n=\omega_n-\Omega$ is the detuning and resonant frequency, and $\gamma_n$ is the total linewidth of the $n^{\text{th}}$ substrate mode. We fit the full multi-parameter RF transmission, $S_{21}^\text{RF}$, using an iterative procedure with particle swarm optimization (PSO) to obtain the substrate mode parameters. These parameters are given in \hyperref[tab:et1]{Extended~Data~Table~1}. Additional details on the fitting procedure and modeling of the substrate modes are presented in the Supplementary Information.

\subsection*{Inferring the electro-optic coupling rate $g_0$}
\label{eocoupling}
We infer the zero-point electro-optic coupling rate using two separate analyses. The first approach uses the measured transduction efficiency, and the second uses the measured optical sideband ratio. Additionally, as discussed in the previous section, coupling to substrate modes eventually impacts the sub-THz intracavity population by adding an additional shift to the sub-THz detuning and total linewidth. Thus, the on-chip efficiency of the transduction process and the sideband ratio are modified (when compared with input-output theory),
\begin{align}
    \eta_\text{OE} &\approx g_0^2 \cdot \left( \frac{\kappa_{e,+}}{\Delta_+^2 + \kappa_+^2/4} \right)\cdot\left( \frac{\kappa_{e,\text{RF}}/2}{(\Delta_\text{RF}+\tilde\delta)^2 + (\kappa_\text{RF}+\tilde\gamma)^2/4} \right) \cdot n_{c,0},\\
    \Upsilon_{+,0} &\approx g_0^2\cdot \Big(\frac{\omega_p+\Omega}{\omega_p}\Big)\cdot\Big(\frac{\kappa_{e,+}}{\Delta_+^2+\kappa_+^2/4}\Big)\cdot\Big(\frac{\kappa_{e,0}/(\Delta_0^2+\kappa_0^2/4)}{1 - 4\kappa_{e,0}\left(\kappa_0-\kappa_\text{e,0}\right)/(4\Delta_0^2 + \kappa_0^2)}\Big)\cdot n_{c,\text{RF}},
\end{align}
where, $n_{c,\text{RF}} = (\kappa_{e,\text{RF}}/2)/((\Delta_\text{RF}+\tilde\delta)^2 + (\kappa_\text{RF}+\tilde\gamma)^2/4) \cdot \frac{P_\text{RF}}{\hbar\Omega}$ and $n_{c,0} = \kappa_{\text{e},0}/(\Delta_0^2 + \kappa_0^2/4)\cdot\frac{P_{0}}{\hbar\omega_p}$. Notice that deducing the electro-optic coupling rate from transduction efficiency is sensitive to both the optical and RF path losses and efficiencies, whereas the sideband ratio is sensitive only to RF path loss and efficiencies because the optical path loss normalizes out in this case, as discussed in the main text. 

To take stock of all measurements, we provide a brief description of how each value is determined in Equations M7 and M8 (see Supplementary Information for a more detailed discussion of this procedure). Firstly, all optical $\kappa$'s and $\kappa_\text{e}$'s are inferred through the self-heterodyne measurement described in the main text. The optical pump detuning $\Delta_0$ is measured as described in~\hyperref[sec:lock]{Methods: Pump detuning calibration and locking}. The blue sideband detuning can be written as, $\Delta_+ = \text{FSR}_+ + \Delta_0 + \Omega$. We write it in this form because in our optical spectroscopy, differences in frequency are more accurately calibrated than the absolute optical wavelength. Note that the quantity $\text{FSR}_+$ is the frequency difference between the pump mode and the blue sideband mode, obtained via the self-heterodyne measurement described above. Lastly, the optical path losses and photodetector power are measured so that we can infer the on-chip photon flux, and thus compute $n_{c,0}$, assuming the input and output grating coupler losses are equal.

For all RF modal parameters, $\Delta_\text{RF}, \tilde\delta,\kappa_\text{RF}, \tilde\gamma,$ and $\kappa_{e,\text{RF}}$ are determined from the multi-parameter fit, to include the effects of the hybridized substrate modes. To determine the RF path loss, we perform a cryogenic calibration procedure (see Supplementary Information)~\cite{multani:2024:quantumlimits1}. This tells us the RF power incident on the chip, and thus provides an estimate of $n_{c,\text{RF}}$ required for Equation M8.

As stated in the main text, we observe that the estimate for $g_0$ from both Equations M7 and M8 converge; providing evidence that the path calibrations and estimates for the mode parameters are self-consistent.

% %%==================================%%
% %%           BACKMATTER             %%
% %%==================================%%
\backmatter
\bmhead{Acknowledgments} We acknowledge Wentao Jiang and Felix Mayor for their assistance in Red Pitaya setup and fiber gluing, Hubert Stokowski and Timothy McKenna for useful discussions on electro-optic theory and modeling, Sandesh Kalantre, Devin Dean, Matthew Maksymowych, Erik Szakiel, Luke Qi, and Samuel Gyger for their help in NbTiN sputtering, and Monika Schleier-Smith and Paul Welander for their experimental support. Lastly, we acknowledge Debadri Das and Samuel Gyger for insightful discussions. This work was supported by the U.S. Army Research Office (ARO)/Laboratory for Physical Sciences (LPS) Modular Quantum Gates (ModQ) program (Grant No. W911NF-23-1-0254), by the Air Force Office of Scientific Research and the Office of Naval Research under award number FA9550-23-1-0338 and the US Department of Energy through grant no. DE-AC02-76SF00515 and via the Q-NEXT Center. This work was also funded by Amazon Web Services.
K.K.S.M. gratefully acknowledges support from the Natural Sciences and Engineering Research Council of Canada (NSERC). J.F.H gratefully acknowledges support from the NSF GRFP under grant number DGE-1656518. Part of this work was performed at the Stanford Nano Shared Facilities (SNSF) and Stanford Nanofabrication Facility (SNF), supported by the National Science Foundation under award ECCS-2026822. Supported in part by the Department of Energy Contract No. DEAC02-76SF00515.

\bmhead{Data Availability}
All data used in this study are available from the corresponding author upon reasonable request.

\bmhead{Competing Interests}
A.H.S.-N. is an Amazon Scholar. The other authors declare no competing interests.

%% EXTENDED DATA FIGURES / TABLES %%
\clearpage

\setcounter{table}{0}
\renewcommand{\tablename}{Extended Table}

\begin{table}[h]\label{tab:et1}
    \centering
    \renewcommand{\arraystretch}{1.25}
    {\rowcolors{2}{}{lightgray!20}
    \begin{tabular}[]{m{3.5cm}w{c}{3.5cm}w{c}{3cm}w{c}{4cm}}
    \toprule
    Cavity Mode  & Frequency $\omega_k/2\pi$ (THz)& Total Loss $\kappa_k/2\pi$ (GHz)& Extrinsic Loss $\kappa_{\mathrm{e},k}/2\pi$ (MHz) \\ \midrule
    
         $\omega_0$ & $194.932$ & $0.201$  &  $88.1$ \\
         $\omega_+$ & $195.037$& $0.210$  &  $87.3$ \\ 
         $\omega_-$ & $194.827$ & $0.199$  & $85.8$ \\ 
         $\omega_\text{RF}$ & $0.106993$ & $1.674$ & $419$
        
         % reset table row counter for coloring
         \setcounter{rownum}{1}
         \\ \addlinespace[20pt]
         \end{tabular}

        % subtable of substrate modes
        \begin{tabular}[]{m{3.5cm}w{c}{3.5cm}w{c}{3cm}w{c}{4cm}}
         \toprule 
         Substrate Mode & Frequency $\omega_n/2\pi$ (GHz)& Total Loss $\kappa_n/2\pi$ (MHz) & Coupling $J_n/2\pi$ (MHz) \\ \hline
         $\omega_\text{1}$ & $102.163$& $302.06$ & 779 \\
         $\omega_\text{2}$ & $102.579$ & $260.11$  & 235 \\
         $\omega_\text{3}$ & $103.746$ & $189.03$  & 0.63 \\
         $\omega_\text{4}$ & $104.978$ & $160.01$  & 682 \\ 
         $\omega_\text{5}$ & $105.843$ & $365.68$  & 857 \\
         $\omega_\text{6}$ & $107.022$ & $445.42$  & 455 \\
         $\omega_\text{7}$ & $108.180$ & $122.75$  & 440 \\
         $\omega_\text{8}$ & $109.784$ & $182.93$  & 467
         
         % reset table row counter for coloring
         \setcounter{rownum}{1}
         \\ \addlinespace[20pt]
         \end{tabular}
         
        \renewcommand{\arraystretch}{1.25}

        \begin{tabular}{m{6.5cm}w{c}{2.5cm}w{c}{2.5cm}w{c}{2.5cm}}
         \toprule
         Transduction Parameter & Symbol & Value & Unit\\ \hline
         EO coupling rate (inferred from efficiency) & $g_0^{\eta}/(2\pi)$ & 707 & Hz  \\
         EO coupling rate (inferred from side band ratio) & $g_0^{\Upsilon}/(2\pi)$ &  685 & Hz \\
        Maximum measured efficiency (on-chip) & $\eta_\text{OE}^{\text{max}}$ & $0.82\times10^{-6}$ & 1 \\
        Single-photon cooperativity & $C_0$ & $6\times10^{-12}$ & 1\\
        Maximum measured cooperativity & $C$ & $1\times10^{-5}$ & 1 \\ 
    \addlinespace[20pt]
    \end{tabular}}
    \caption{Device parameters as inferred from independent measurements, described in the Results section. The optical measurements are taken when no RF power was applied. The sub-THz substrate mode parameters are identified from a fit procedure discussed in supplementary information. The $J_n$ values reported here, corresponding to coupling between the substrate mode and the sub-THz resonator mode at base temperature (\qty{4.9}{\K}). Note that we cannot infer the substrate mode $\omega_\text{RF,3}$ coupling to the sub-THz resonator at base temperature. We attribute this to only a very weak coupling between these two at base temperature that is not extracted during fitting.}
    \label{tab:extended_mode_params}
\end{table}
\clearpage

% Change figure name format
\setcounter{figure}{0}
\renewcommand{\figurename}{Extended Fig.}

%% EF1
\clearpage
\begin{figure*}[b]
    \centering
    \includegraphics[width=\linewidth]{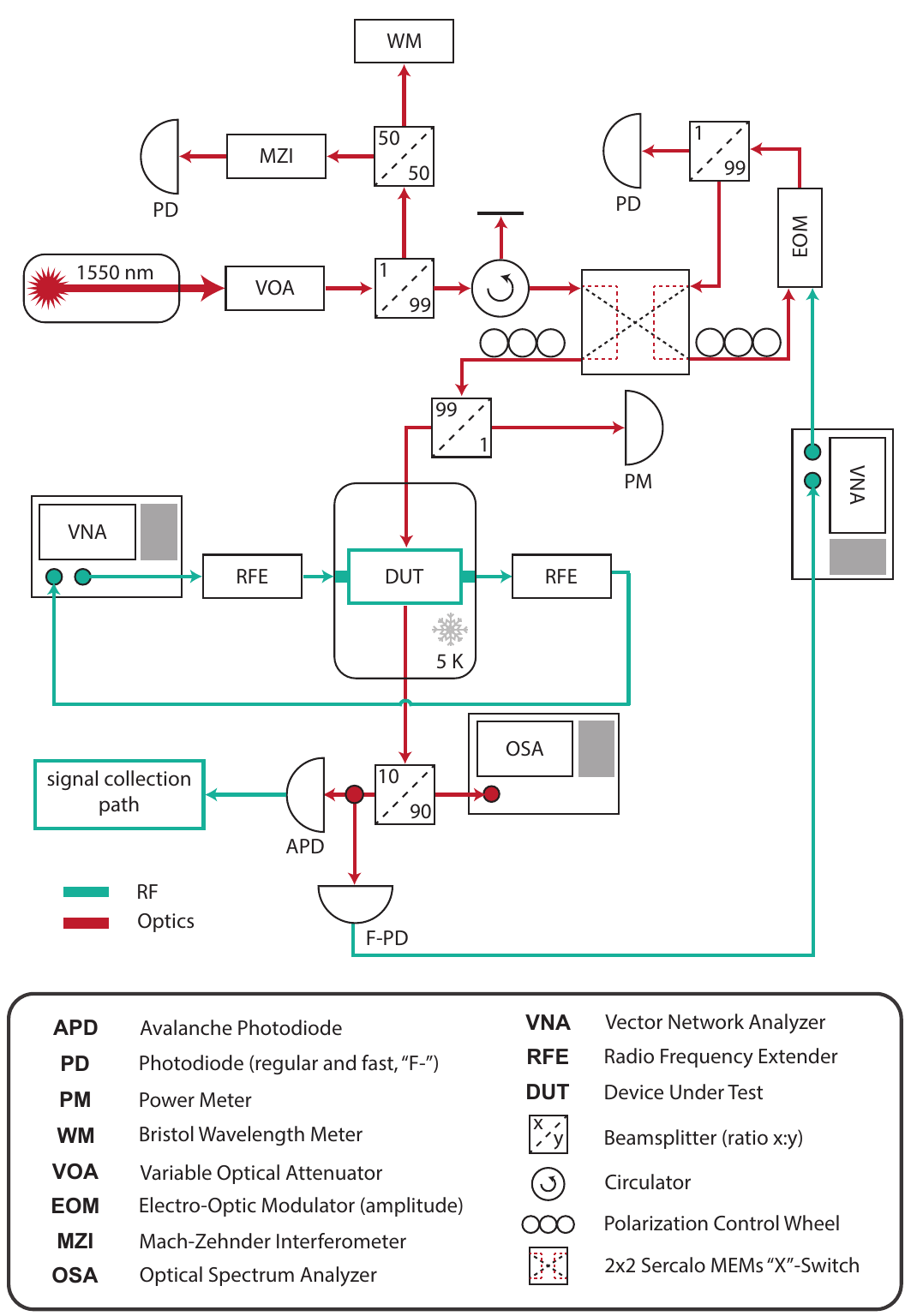}
    \caption{Complete measurement schematic depicting all paths and equipment used for the experiment. Red arrows denote optical paths, while aqua arrows denote RF paths. The device under test (DUT) is held within a cryostat at $T\approx \qty{4.9}{\K}$.}
    \label{fig:ef1}
\end{figure*}

%% EF2
\clearpage
\begin{figure*}[b]
    \centering
    \includegraphics[width=\linewidth]{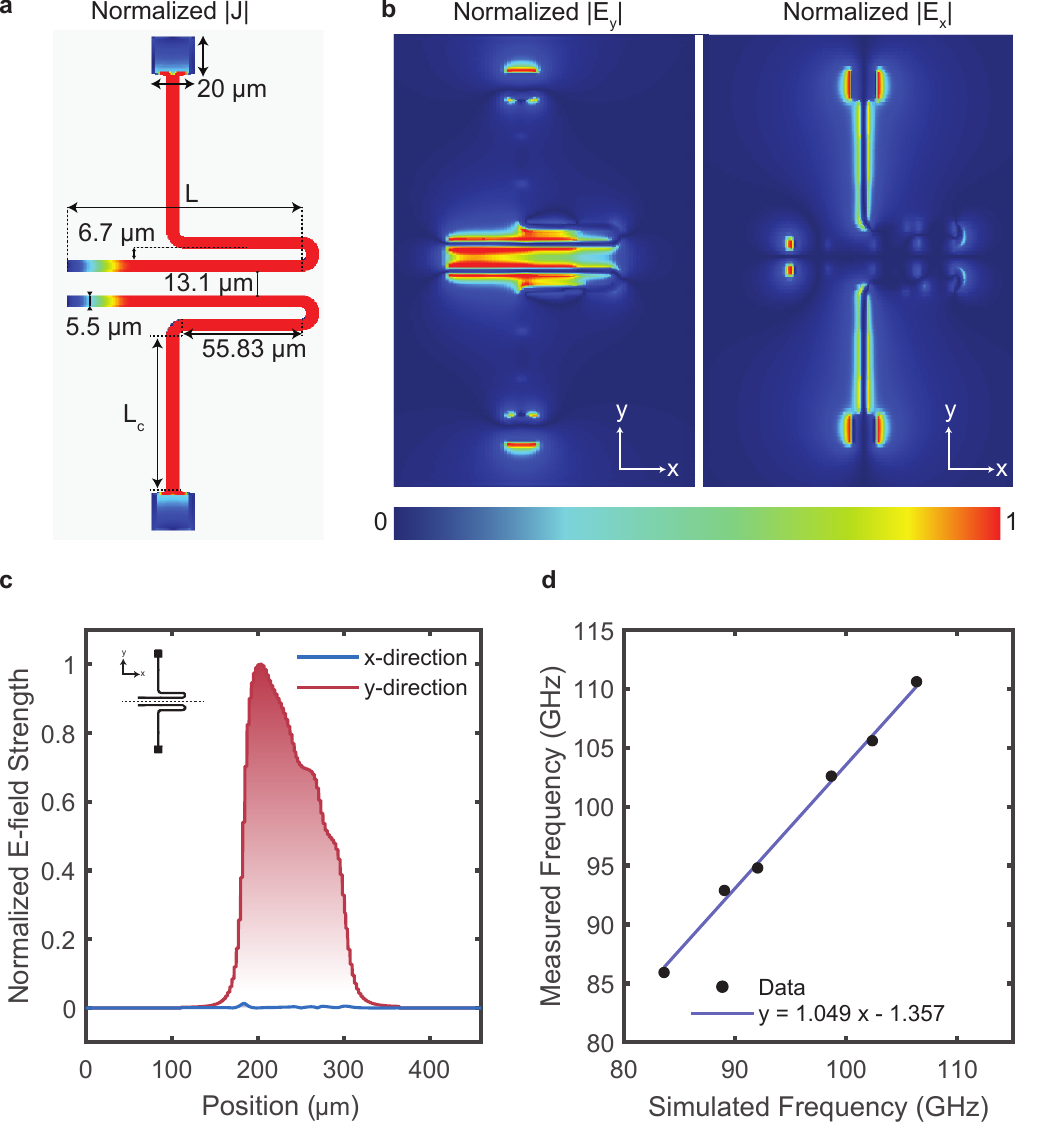}
    \caption{Sub-THz resonator design. \textbf{a} SONNET-simulated normalized current density and geometry of the sub-THz electrodes of the device shown in the main text (on resonance). The device in the main text has $L=\qty{111.65}{\um}$ and $L_c=\qty{82}{\um}$. \textbf{b} SONNET simulation of the normalized electric field magnitudes in the $x$ and $y$ direction of the sub-THz electrodes (on resonance). \textbf{c} A line-cut of the data in sub-figure (b), showing the electric field distribution along the line-cut. The location of the line-cut is shown in the inset on the top-left. \textbf{d} Data of simulations and measurements of the NbTiN-witness resonators. We used this data to design the NbTiN electrode geometry of the device in the main text. To vary the frequency, we only changed the length of the straight section $L$.}
    \label{fig:ef2}
\end{figure*}

%% EF3
\clearpage
\begin{figure*}[b]
    \centering
    \includegraphics[width=\linewidth]{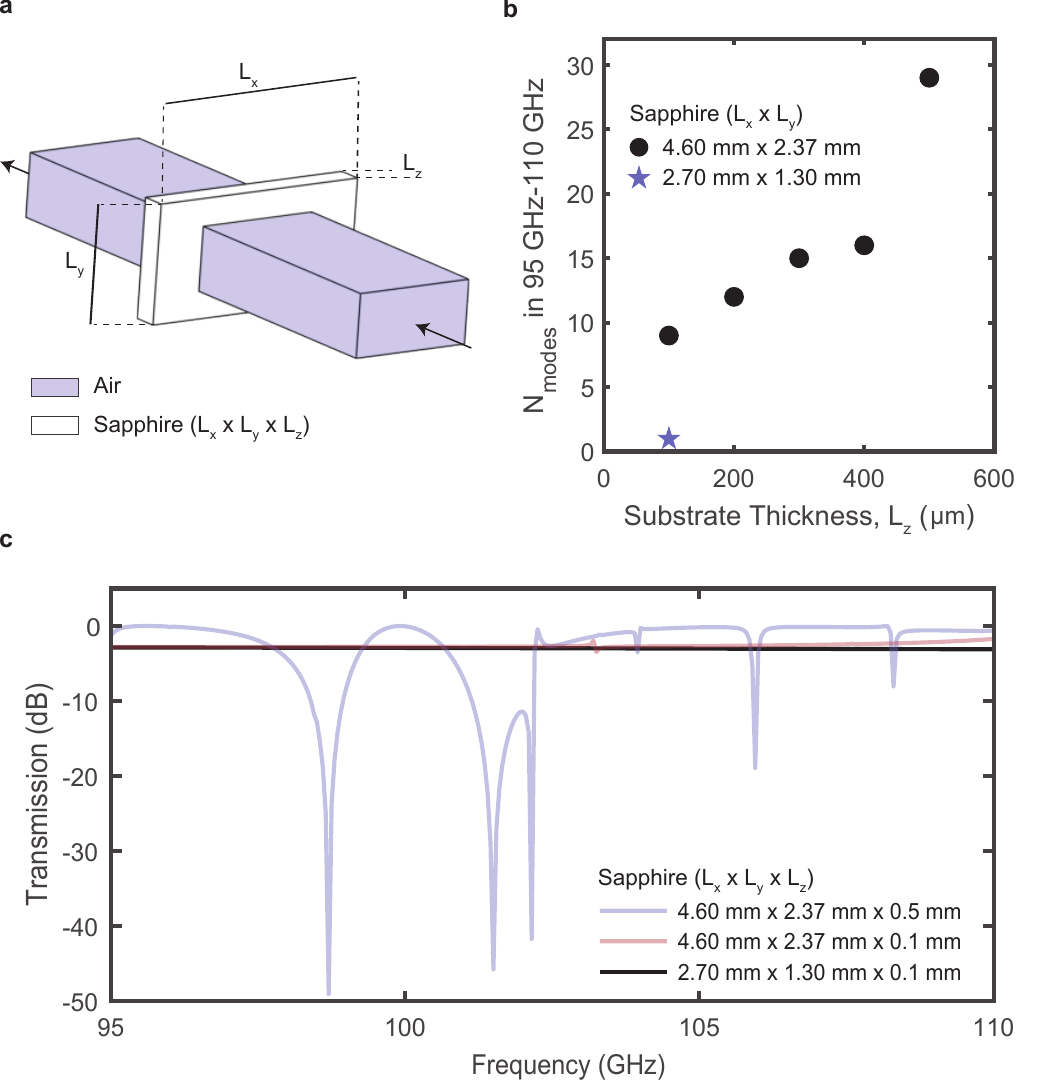}
    \caption{Simulations of substrate mode behavior as a function of chip geometry.~\textbf{a} Labeled COMSOL geometry of sapphire substrate in a WR10 waveguide. The input and output waveguide ports are labeled by arrows, and all boundaries are set to be perfect electrical conductors.~\textbf{b} Number of modes supported by the substrate in the geometry defined in subfigure (a) within the frequency range of interest. The number of modes was given by COMSOL via eigenfrequency simulations. Black dots indicate a substrate with the same dimensions as our sample (top-right-most datapoint), and the reduction in modes as the substrate thickness is reduced. The blue star indicates a dimensionality that should support minimal substrate modes in our target frequency range.~\textbf{c} COMSOL simulations of the transmission spectrum through the structure in subfigure (a), revealing resonances supported by the Sa substrate. Reducing the dimensionality drastically suppresses the number of supported substrate modes in our target frequency range in this coupling geometry.}
    \label{fig:ef3}
\end{figure*}

%% EF 4
\clearpage
\begin{figure*}[b]
    \centering
    \includegraphics[]{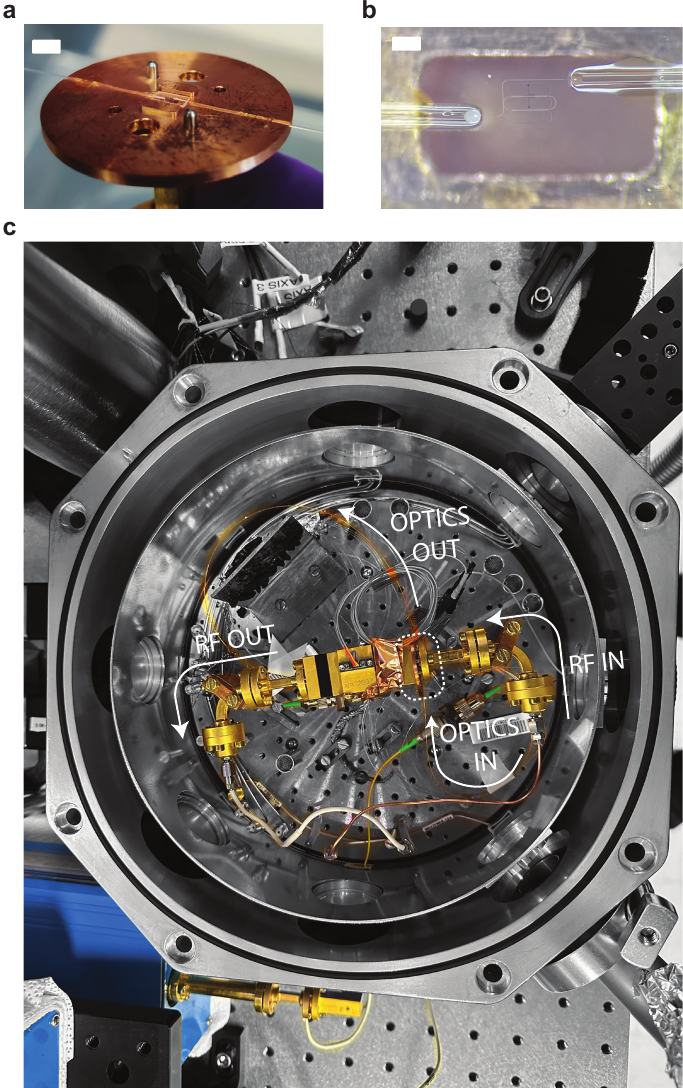}
    \caption{Images of the packaging and cryostat setup. \textbf{a} Smartphone image of the package device inside of the copper holder. \hyperref[fig:fig1]{Figure~1b} in the main text is inspired by this image. The scale bar represents \qty{5}{\mm}. \textbf{b} Microscope image of the device immediately after the angle cleaved fibers were UV-cured onto the chip. The scale bar represents \qty{300}{\um}. \textbf{c} Color-masked smartphone image of the cryostat preceding the cooldown. The regions with color emphasize the RF signal path and the optical signal path. The dashed box highlights where the device is fixed. Block diagrams of the RF and optical paths are attached in the Supplementary Information.}
    \label{fig:ef4}
\end{figure*}
%TC:endignore
\clearpage

%%%%% UPDATED COMMANDS %%%%%%%%
\section*{Supplementary Information}

\renewcommand{\thesection}{S\arabic{section}}
\renewcommand{\figurename}{Supplementary Fig.}
\renewcommand{\theequation}{S\arabic{equation}}
\setcounter{equation}{0}
\setcounter{figure}{0}
\setcounter{secnumdepth}{3}
\setcounter{page}{0}
\thispagestyle{empty}
\tableofcontents
%%%%%%%%%%%%%%%%%%%%%%%%%%%%%%%%%%%%%%%%%%%%%%%%% SECTION
\section{Device physics and theory}
\label{devicetheory}
In this section, we describe the modeling of our device. We begin with the Hamiltonian first and solve the resulting equations of motion, derive an expression for the electro-optic interaction rate and discuss the electrode design, introduce a phenomenological model of the interaction between the substrate modes and the sub-THz cavity, and lastly describe the impact the substrate modes on the device's nominal performance. 

%%%%%%%%%%%%%%%%%%%%%%%%%%%%%%%%%%%%%%%% SUBSECTION
\subsection{Nominal solutions of the equations of motion and practical considerations}
\label{nominalsol}
We begin with the Hamiltonian of our cavity electro-optic system under the rotating wave approximation,
\begin{equation}
    \hat{H}/\hbar=\hat{H}_0/\hbar+g_0\Big(\hat{a}^{+\dagger} \hat{a}^0\hat{b}+\text{h.c.}\Big)+g_0\Big(\hat{a}^{-\dagger} \hat{a}^0 \hat{b}^\dagger + \text{h.c.}  \Big),
\end{equation}
where,
\begin{equation}
    \hat{H}_0/\hbar = \omega_\text{RF}\hat{b}^\dagger \hat{b}+\omega_-\hat{a}^{-\dagger} \hat{a}^- + \omega_0 \hat{a}^{0\dagger} \hat{a}^0 + \omega_+ \hat{a}^{+\dagger} \hat{a}^+.
\end{equation}
Here $\hat{a}_+$, $a_0$, and $\hat{a}_-$ are the annihilation operators for the blue ($\omega_+$), pump ($\omega_0$), and red ($\omega_-$) optical modes respectively; $b$ is the annihilation operator for the sub-THz ($\omega_\text{RF}$) mode; $g_0$ is the single-photon electro-optic coupling rate. In our experiment, all fields are coherent, so we displace each operator and neglect the fluctuations. Thus, we write the coupled mode equations of the classical amplitudes denoted by $\alpha$'s for the intracavity optical fields and $\beta$ for the sub-THz field below. Note in our experiments, we drive the sub-THz mode and the optical pump  mode so we include input fields for the optical pump and the sub-THz modes, $|\alpha_\text{in}|^2 = P_0/(\hbar\omega_p)$, and $|\beta_\text{in}|^2 = P_\text{RF}/(\hbar\Omega)$,

\begin{align}
\dot\alpha_0 & =-(i\Delta_0+\kappa_0/2)\alpha_0 - \sqrt{\kappa_{e,0}}\alpha_\text{in}\\
\dot\alpha_+ &= -(i\Delta_++\kappa_+/2)\alpha_+-ig_0\alpha_0\beta\\
\dot\alpha_- &= -(i\Delta_-+\kappa_-/2)\alpha_--ig_0\alpha_0\beta^\ast\\
\dot\beta &= -(i\Delta_\text{RF}+\kappa_\text{RF}/2)\beta - ig_0\alpha_+\alpha_0^\ast -ig_0 \alpha_-^\ast \alpha_0-\sqrt{\kappa_{e,\text{RF}}/2}\cdot\beta_\text{in},
\end{align}
where $\Delta_\text{RF}=\omega_\text{RF}-\Omega$, $\Delta_0= \omega_0-\omega_p$, $\Delta_+=\omega_+-(\omega_p+\Omega)$, and $\Delta_-=\omega_--(\omega_p-\Omega)$; $\omega_p$ and $\Omega$ are the optical and sub-THz drive frequencies, respectively. Additionally, the sub-THz mode is modeled as being double-side coupled, which is why we have a factor of 1/2 for the input field coupling rate. Thus, the total linewidth is defined as $\kappa = \kappa_i + \kappa_e$ for all modes. In the steady-state, the intracavity field amplitudes are,
\begin{align}
\alpha_0&=\frac{-\sqrt{\kappa_{e,0}}}{i\Delta_0+\kappa_0/2}\cdot \alpha_\text{in}\\
\beta&= \frac{-\sqrt{\kappa_{e,\text{RF}}/2}}{i\Delta_\text{RF}+\kappa_\text{RF}/2+\cancel{g_0^2|\alpha_0|^2\Big(\frac{1}{i\Delta_++\kappa_+/2}+\frac{1}{i\Delta_--\kappa_-/2}\Big)}}\cdot \beta_\text{in}\\
\alpha_+&=\frac{-ig_0}{i\Delta_++\kappa_+/2}\cdot\alpha_0\beta\\
\alpha_-&=\frac{-ig_0}{i\Delta_-+\kappa_-/2}\cdot\alpha_0\beta^\ast,
\end{align}
where the diagonal strike-out denotes that we assume we are in the low-cooperativity limit. This is equivalent to assuming that there is no back-action on the optical modes from the dynamics of the sub-THz mode. Using the input-output boundary condition, $\alpha_k^\text{out} = \alpha_k^\text{in} + \sqrt{\kappa_{e,\text{k}}}\cdot\alpha_k$, and similarly $\beta^\text{out} = \beta^\text{in} + \sqrt{\kappa_{e,\text{RF}}/2}\cdot\beta$, we can compute the output fields,

\begin{align}
\alpha_0^\text{out}&=\Big(1-\frac{\kappa_{e,0}}{i\Delta_0+\kappa_0/2}\Big)\cdot \alpha_\text{in}\\
\beta^\text{out}&= \Big(1-\frac{\kappa_{e,\text{RF}}/2}{i\Delta_\text{RF}+\kappa_\text{RF}/2}\Big)\cdot\beta_\text{in}\\
\alpha_+^\text{out}&=\frac{ig_0\sqrt{\kappa_{e,+}}}{i\Delta_++\kappa_+/2}\cdot\frac{\sqrt{\kappa_{e,0}}}{i\Delta_0+\kappa_0/2}\cdot\alpha_\text{in}\cdot\beta\\
\alpha_-^\text{out}&=\frac{ig_0\sqrt{\kappa_{e,-}}}{i\Delta_-+\kappa_-/2}\cdot\frac{\sqrt{\kappa_{e,0}}}{i\Delta_0+\kappa_0/2}\cdot\alpha_\text{in}\cdot\beta^\ast.
\end{align}

With the solutions to the coupled mode equations, we compute two quantities that we use to estimate the $g_0$ of our device: the efficiency and the side-band ratio. We begin with the device efficiency in the low cooperativity limit, 
\begin{align}
    \eta_\text{OE} &= \left|\frac{\alpha_+^\text{out}}{\beta_\text{in}}\right|^2\\
    &\approx g_0^2 \cdot \left( \frac{\kappa_{e,+}}{\Delta_+^2 + \kappa_+^2/4} \right)\left( \frac{\kappa_{e,\text{RF}}/2}{\Delta_\text{RF}^2 + \kappa_\text{RF}^2/4} \right) \cdot n_{c,0} 
\end{align}
where $n_{c,0} = |\alpha_0|^2 = \kappa_{e,0}/(\Delta_0^2 + \kappa_0^2/4) |\alpha_\text{in}|^2$ is the intracavity pump photon number. If we assume that $\omega_+ = \omega_0 + \omega_\text{RF}$ and we drive on resonance $\omega_p=\omega_0$ with $\Omega = \omega_\text{RF}$, then the expression simplifies to
\begin{align}
    \eta_\text{OE} &\approx 4 C_0 \left(\frac{\kappa_{e,+}}{\kappa_+}\right)\left(\frac{\kappa_{e,\text{RF}}}{2\kappa_\text{RF}}\right) \cdot n_{c,0} 
\end{align}
Here $C_0 \equiv 4g_0^2/(\kappa_+\kappa_\text{RF})$ is the single-photon cooperativity.

The second quantity comes from the output sideband power ratio,
\begin{align}
    \Upsilon_{+,0} &= \frac{P_+^\text{out}}{{P_0^\text{out}}}=\frac{\omega_p+\Omega}{\omega_p}\left|\frac{\alpha_+^\text{out}}{\alpha_0^\text{out}}\right|^2\\
    &\approx g_0^2\cdot \Big(\frac{\omega_p+\Omega}{\omega_p}\Big)\cdot\frac{\Big(\frac{\kappa_{e,+}}{\Delta_+^2+\kappa_+^2/4}\Big)\Big(\frac{\kappa_{e,0}}{\Delta_0^2+\kappa_0^2/4}\Big)}{1 - 4\frac{\kappa_{e,0}\left(\kappa_0-\kappa_\text{e,0}\right)}{4\Delta_0^2 + \kappa_0^2}}\cdot n_{c,\text{RF}},
\end{align}
where $n_{c,\text{RF}} = |\beta|^2 = (\kappa_{e,\text{RF}}/2)/(\Delta_\text{RF}^2 + \kappa_\text{RF}^2/4) |\beta_\text{in}|^2$ is the intracavity sub-THz photon number. To simplify, we can assume the same as above (on-resonance), to arrive at the expression,
\begin{align}
    \Upsilon_{+,0} &\approx 4C_0 \cdot \frac{\eta_{e,+}\eta_{e,0}}{(1-2\eta_{e,0})^2}\cdot \frac{\omega_+}{\omega_0} \cdot n_{c,\text{RF}},
\end{align}
given that $\eta_{e}=\kappa_e/\kappa$ for each of the optical modes. Note, we can also compute the sideband ratio between the red and blue sidebands,
\begin{align}
    \Upsilon_{+,-} &= \frac{\eta_{e,+}}{\eta_{e,-}}\cdot\frac{Q_+}{Q_-}.
\end{align}
This equation predicts that if the red and blue modes have different quality factors that the observed peak power will be different. Indeed in Figure 3 of the main text, we observe the red and blue sidebands differ in the measured power, consistent with the above expression.

In our experiment, we also include the impacts of the substrate modes together with the above expressions of $\eta_\text{OE}$ and $\Upsilon_{+,0}$ to compute an estimate of the electro-optic coupling rate $g_0$ (see~\hyperref[impactnominal]{Supplementary Section S1.4}). Calibration of optical and sub-THz path losses is important in computing both the on-chip efficiency $\eta_\text{OE}$ and the sideband ratio $\Upsilon_{+,0}$. In the case of efficiency, we measure the output photon flux ($|\alpha_+^\text{out}|^2$) via the optical spectrum analyzer and infer the incident sub-THz ($|\beta_\text{in}|^2$) and intracavity optical photon number ($n_{c,0}$) through our calibrations. Therefore, the measured efficiency and on-chip efficiency are related by optical and sub-THz path losses,
\begin{align}
    \eta_\text{OE}^\text{meas} = \eta_\text{output}^\text{opt}\eta_\text{input}^\text{RF}\cdot\eta_\text{OE},
\end{align}
where the output optical efficiency includes chip-fiber loss and path loss from the fiber inside of the cryostat to the optical spectrum analyzer, and the input sub-THz efficiency includes the respective path loss to the device. The sub-THz input efficiency includes the path loss of the sub-THz source from outside of the cryostat to right before the device and the loss reduction that comes from the increased conductivity of the waveguides inside the cryostat due to low-temperature operation. Importantly, $\eta_\mathrm{OE}$ also depends on the intracavity photon number in the optical pump mode, which is in part calculated from the input optical pump power and the input loss,  $\eta_\text{input}^\text{opt}$. The optical input loss includes the path loss from the fiber directly outside of the cryostat to the fiber right before the fiber-chip interface and the fiber-chip loss. In the main text, we report $\eta_\text{OE}$.

In the case of the sideband ratio, the optical input and output efficiencies do not enter the equation. This is because we directly measure the output power of the pump and the converted optical power on the optical spectrum analyzer; and both of these fields travel the same optical path, so path efficiencies are canceled in the ratio. Therefore, this is an optically self-calibrating measurement. Thus, the measured sideband ratio depends only on the sub-THz path loss,

\begin{align}
    \Upsilon_{+,0}^\text{meas} &= \eta_\text{input}^{\text{RF}} \cdot \Upsilon_{+,0}.
\end{align}

Further details of our calculations and calibrations regarding device performance can be found in the Methods section of the main text, \hyperref[impactnominal]{Supplementary Section S1.4}, and \hyperref[couplingfit]{Supplementary Section S2}.

%%%%%%%%%%%%%%%%%%%%%%%%%%%%%%%%%%%%%%%% SUBSECTION

\subsection{The electro-optic coupling rate and sub-THz electrode design considerations}
In the main text, we explain that the sub-THz superconducting resonator's electric field distribution must be unipolar across the TFLN interaction region. Here we present a brief discussion of the reason behind this, in addition to a derivation of the coupling rate.

\subsubsection{Deriving an expression for $g_0$}
In systems similar to ours, the electro-optic coupling rate ($g_0$) can be derived from the nonlinear interaction energy arising from the $\chi^{(2)}$ nonlinearity,

\begin{equation}
\begin{split}
    \mathcal{H}_{\text{NL}}^{(2)} \to \mathcal{\hat H}_{\text{NL}}^{(2)} &=\frac{2\epsilon_0}{3}\mathbf{\hat E}\cdot\Big(\chi^{(2)}:\mathbf{ \hat E}\otimes \mathbf{ \hat E}\Big)\\
    &= \frac{2}{3}\epsilon_0\sum_{k=1}^{3}\sum_{\ell=1}^{3}\sum_{m=1}^{3}\chi^{(2)}_{k\ell m}\hat{E}_k\hat{E}_\ell\hat{E}_m.
    \end{split}
\end{equation}

Expanding the electric field in terms of operators, $\mathbf{\hat E}(\mathbf{r},t) = \sum_{\alpha=1}^{4} \mathbf{E}^\alpha(\mathbf{r})\hat{a}_\alpha e^{-i(\omega_\alpha t-k_\alpha z)}+\text{h.c.}$ The indices represent the four modes of interest with bosonic annihilation operators, $\{\hat{a}_{-},\hat{a}_{0},\hat{a}_{+}, \hat{b}\}\equiv\{\hat{a}_1, \hat{a}_2,\hat{a}_3,\hat{a}_4\}$. Note that the component subscript $i$ in, $E_i^{\alpha}(\mathbf{r})$, denotes the coordinate axes, $1\mapsto x, 2\mapsto y, 3\mapsto z$. Writing out the tensor product and keeping only energy conserving terms (terms that obey the condition $|\omega_\pm-\omega_0|=\omega_\text{RF}$, or, using the $\alpha$-indices,  $|\omega_{1,3}-\omega_2|=\omega_4$), we obtain the expression
\begin{align*}
\hat{E}_k\hat{E}_\ell\hat{E}_m &= \Big[\hat{a}_+^{\dagger}\hat{a}_{0}\hat{b}\Big(E_k^{+\ast}E_\ell^{0}E_m^\text{RF} + E_\ell^{+\ast}E_m^{0}E_k^\text{RF} + E_m^{+\ast}E_k^{0}E_\ell^\text{RF} \Big)e^{i(k_0+k_\text{RF}-k_+)x} + \text{h.c.}\Big]
\\
&+ \Big[\hat{a}_-^{\dagger}\hat{a}_{0}\hat{b}^{\dagger}\Big(E_k^{-\ast}E_\ell^{0}E_m^{\text{RF}^\ast} + E_\ell^{-\ast}E_m^{0}E_k^{\text{RF}^\ast} +E_m^{-\ast}E_k^{0}E_\ell^{\text{RF}^\ast} \Big)e^{i(k_0-k_\text{RF}-k_-)x} + \text{h.c.}\Big].
\end{align*}
Comparing to Equation S1, and matching coefficients, we obtain an expression for the electro-optic coupling rate,

\begin{equation}
    \hbar g_0 = \frac{2}{3}\epsilon_0N_{+}N_0N_\text{RF}\iiint_{\mathbb{R}^3}\pmb{\psi}^{+\ast} \cdot \Big(\chi^{(2)} : \pmb{\psi}^0 \otimes\pmb{\psi}^\mathrm{RF}\Big) e^{i(k_0+k_\text{RF}-k_+)x}\mathrm{d}V.
\end{equation}
In the expression above, we have introduced unitless normalized spatial distributions, $\mathbf{E}^\alpha = N_\alpha \pmb{\psi}^\alpha$. This normalization ensures if there is one photon in the mode, that the electric field energy is equal to $\hbar\omega_\alpha/2$. We can write the normalization constants in terms of the field mode volumes $V_{\alpha}$ as,

\begin{equation}
    N_\alpha = \sqrt{\frac{\hbar\omega_\alpha}{2\epsilon_0V_\alpha}} = \sqrt{\frac{\hbar\omega_\alpha}{2\epsilon_0}}\times\sqrt{\frac{1}{\iiint \pmb{\psi}^{\alpha\ast} \cdot(\epsilon_r\pmb{\psi}^\alpha) \mathrm{d}V}}.
\end{equation}

\subsubsection{Superconducting electrode design considerations for longitudinally varying fields}
A large part of the engineering that goes into optimizing $g_0$ is the placement of the electrode with respect to the optical waveguides. In our device, the dominant nonlinearity is along the  crystal axis ($\hat{\pmb{z}}$-axis), corresponding to the nonlinearity $\chi^{(2)}_{333}$. We place electrodes (total length $L_e$) across one section of the optical racetrack resonator (total length $L_o$) to drive the electro-optic interaction. If we separate the RF field and the integral into longitudinal and transverse components, we can reveal the important design considerations, i.e. ${\psi}_i^\mathrm{RF} = \psi^{t,\mathrm{RF}}_i(z,y)  \psi^{\ell,\mathrm{RF}}_i(x)$ and $\iiint \mathrm{d}V = \iint\mathrm{d}A \int \mathrm{d}x$,
\begin{equation}
    \hbar g_0 \approx 2\epsilon_0N_+N_0N_\mathrm{RF} \times \Big(\iint_{A} \chi^{(2)}_{333}|\psi_3^0|^2 \psi^{t,\mathrm{RF}}_3\mathrm{d}A\Big) \times \Big(\int_0^{L_e}  \psi^{\ell,\mathrm{RF}}_3(x) e^{i(k_0+k_\text{RF}-k_+)x} \mathrm{d}x\Big).
\end{equation}
Equation S27 has three terms: (i) constants proportional to $1/\sqrt{V_\alpha}$ ($V_\alpha$ are the mode volumes of each field), (ii) an overlap integral of the transverse fields (we assume the two optical field distributions corresponding to the pump mode $\omega_0$ and the sideband mode $\omega_+$ are equivalent in our device) within the cross section of the TFLN, and (iii) an integration along the length of modulation, which includes a phase matching term. Because our sub-THz resonator is not a lumped-element , we need to consider the phase matching term carefully. In our case, the electrodes are much shorter than the optical racetrack, and we observe $(k_0-k_+)x\approx 2\pi/L_ox \ll 1$ for $x \in [0,L_e]$. That is, the phase mismatch of neighboring optical modes in a racetrack resonator is minimal compared to the phase mismatch that arises from the longitudinally varying RF field. 

Now, consider the simplest electrode design: two metal slabs, one placed above and one placed below the TFLN waveguide, in a coplanar stripline geometry. The amplitude of the RF field between these electrodes is $\psi_3^{\ell,\mathrm{RF}} = \Psi$. In this scenario, the electrodes can be open-ended or close-ended on one side i.e. a half-wave or quarter-wave coplanar stripline resonator. In the case of the half-wave resonator the wavenumber for the fundamental mode is $k_\mathrm{RF} = \pi/L_e$, therefore phase matching integral can be written as,
\begin{equation}
    I = \int_0^{L_e/2} \Psi e^{i\pi x/L_e}\mathrm{d}x - \int_{L_e/2}^{L_e} \Psi e^{i\pi (x-L_e/2)/L_{e}}\mathrm{d}x = 0.
\end{equation}
The phase term is split into two regions because at $x=L_e/2$ the polarity of the electric field goes from parallel to antiparallel with respect to the crystal axis. So, in the case of a simple half-wave resonator, the electro-optic coupling rate vanishes. 

For the quarter-wave resonator, the rate does not vanish because the field does not change polarity. However, the electrode would need to cross-over the TFLN waveguide, which introduces loss for the optical modes (due to proximity to the metal) and loss for the RF mode (due to proximity to the oxide cladding atop the optics). Therefore, for the device presented in the main text, we start with a half-wave resonator and modify the electrodes so that the RF field distribution across the TFLN is more approximate to a quarter-wave resonator.

%%%%%%%%%%%%%%%%%%%%%%%%%%%%%%%%%%%%%%%% SUBSECTION
\subsection{Input-output model of the sub-THz resonator in the presence of substrate modes}
\label{substratemodes}
As described in the main text, the sub-THz spectrum contains a cacophony of substrate-supported modes that couple to the sub-THz superconducting resonator. We are able to probe the sub-THz mode via a dipole interaction, just like an antenna. However, we posit that the electric field supported by the resonator can also couple to otherwise inaccessible (dark) modes within the Sapphire substrate. In this section, we derive a model to describe this situation.

Suppose there are $N$ total substrate modes, each with a coupling rate $J_n$ to the sub-THz mode, total linewidth $\gamma_n$, and resonant frequency $\omega_n$. The coupled mode equations of the classical amplitudes in a frame rotating with the sub-THz drive frequency $\Omega$, can be written as,
\begin{align}\label{eq:sub_modes}
    \dot\beta &= -\left(i\Delta_\text{RF}+\frac{\kappa_\text{RF}}{2}\right)\beta - i\sum_{n=1}^{N}J_n^\ast s_n -\sqrt{\frac{\kappa_{e,\text{RF}}}{2}}\beta_\text{in}\\
    \dot s_n &= -\left(i\Delta_n+\frac{\gamma_n}{2}\right)s_n - iJ_n\beta,
\end{align}
where, $s_n$ is the $n^\mathrm{th}$ substrate mode intracavity amplitude, $\Delta_\text{RF} = \omega_\text{RF} - \Omega$, and $\Delta_n = \omega_n - \Omega$. Note the sub-THz resonator is double-side coupled, necessitating the factor of two in the drive term, just as in the above section. In steady-state, the sub-THz intracavity amplitude is,
\begin{align}
 \beta &= \frac{-\sqrt{\kappa_{e,\text{RF}}/2}}{i(\Delta_\text{RF}+\tilde\delta) + (\kappa_\text{RF}+\tilde\gamma)/2 }\cdot\beta_\text{in}.
\end{align}
where,
\begin{align}
    \tilde\delta &= -\sum_{n=1}^N \frac{|J_n|^2\Delta_n}{\Delta_n^2+\gamma_n^2/4} \\
    \tilde\gamma/2 &=  \sum_{n=1}^N \frac{|J_n|^2\gamma_n/2}{\Delta_n^2 + \gamma_n^2/4}.
\end{align}

Using the input-output boundary condition $\beta_\text{out} = \beta_\mathrm{in} + \sqrt{\kappa_{e,\text{RF}}/2}\cdot\beta$, we can write the sub-THz output field amplitude as,
\begin{equation}
    \beta_\mathrm{out} = \underbrace{\left(1- \frac{\kappa_{e,\text{RF}}/2}{i(\Delta_\text{RF}+\tilde \delta) + (\kappa_\text{RF} + \tilde\gamma)/2}\right)}_{\text{transmission coefficient, } S_{21}}\cdot\beta_\mathrm{in}.
\end{equation}

Here we see that coupling to the substrate modes is indistinguishable from a shift in detuning and an increased linewidth of the superconducting sub-THz mode. As a result, the intracavity population of the sub-THz mode can be affected, which impacts the device performance (see next section). Thus, carefully fitting and understanding the substrate mode parameters is needed. We provide details of our fitting methodology in \hyperref[substratefit]{Supplementary Section S3}.

%%%%%%%%%%%%%%%%%%%%%%%%%%%%%%%%%%%%%%%% SUBSECTION
\subsection{Impact of substrate mode hybridization on the electro-optic conversion efficiency and sideband ratio}
\label{impactnominal}
In this section, we discuss the implications of the substrate modes on the nominal device performance as discussed in~\hyperref[nominalsol]{Supplementary Section S1.1}. As shown in the previous section, coupling to substrate modes impacts the sub-THz intracavity population by adding an additional shift to the sub-THz resonance detuning and total linewidth. Thus, the on-chip efficiency of the transduction process and the sideband ratio are modified according to,
\begin{align}
    \eta_\text{OE} &\approx g_0^2 \cdot \left( \frac{\kappa_{e,+}}{\Delta_+^2 + \kappa_+^2/4} \right)\cdot\left( \frac{\kappa_{e,\text{RF}}/2}{(\Delta_\text{RF}+\tilde\delta)^2 + (\kappa_\text{RF}+\tilde\gamma)^2/4} \right) \cdot n_{c,0},\\
    \Upsilon_{+,0} &\approx g_0^2\cdot \Big(\frac{\omega_p+\Omega}{\omega_p}\Big)\cdot\Big(\frac{\kappa_{e,+}}{\Delta_+^2+\kappa_+^2/4}\Big)\cdot\Big(\frac{\frac{\kappa_{e,0}}{\Delta_0^2+\kappa_0^2/4}}{1 - 4\frac{\kappa_{e,0}\left(\kappa_0-\kappa_\text{e,0}\right)}{4\Delta_0^2 + \kappa_0^2}}\Big)\cdot n_{c,\text{RF}},
\end{align}
where,
\begin{align}
    n_{c,\text{RF}} &= |\beta|^2 \\
    &= \frac{\kappa_{e,\text{RF}}/2}{(\Delta_\text{RF}+\tilde\delta)^2 + (\kappa_\text{RF}+\tilde\gamma)^2/4 }\cdot \frac{P_\text{RF}}{\hbar\Omega}.
\end{align}
We use these equations for our estimation of $g_0$ in the main text, with further details on the fitting procedure in~\hyperref[couplingfit]{Supplementary Section S2}.

%%%%%%%%%%%%%%%%%%%%%%%%%%%%%%%%%%%%%%%% SECTION
\section{Electro-optic coupling rate inference and related data analysis methodology}
\label{couplingfit}

In this section, we provide details on the procedures used to estimate the electro-optic coupling rate $g_0$. As discussed in the main text and previous supplementary sections, we infer $g_0$ using two different quantities: the on-chip transduction efficiency (Eq. S35); and the ratio of the converted sideband power and the pump power (Eq. S36).

%%%%%%%%%%%%%%%%%%%%%%%%%%%%%%%%%%%%%%%% SUBSECTION
\subsection{Optical and RF path calibrations}\label{calibrations}
We begin by describing how we calibrate path losses. As mentioned in~\hyperref[nominalsol]{Supplementary Section S1.1}, the measured efficiency includes both optical and RF path losses. For the RF path, we calculate the power incident to the chip,
\begin{align}
    |\beta_\text{in}|^2 = \eta_\text{input}^\text{RF} |\beta_\text{source}^\text{RF}|^2,  
\end{align}
where $|\beta_\text{source}^\text{RF}|^2 = P_\text{source}^\text{RF}/(\hbar \Omega)$. By using a diode-based power detector (Pacific Millimeter-wave Products), we measure the power of the RF extender to be $P_\text{source}^\text{RF} = \qty{1.87}{\milli\watt}$ at $\Omega = 2\pi\cdot\qty{105.285}{\GHz}$. This frequency is associated with the peak transducer response. The same detector is used to measure the insertion loss from the source to the device at room temperature. We measure this path loss to be $-21.2$ dB. Notably, the portion of the RF path that is thermalized to \qty{4}{\K} exhibits reduced loss at cryogenic temperatures, due to increased conductivity of the WR10 waveguides. We measure this cryogenic ``gain'' to be $3.56$ dB. Therefore, the total RF path loss up to the device at cryogenic temperatures is $-17.6$ dB or $\eta_\text{input}^\text{RF} = 1.72\%$. More details on the cryogenic calibration procedure can be found in~\cite{multani:2024:quantumlimits}.

For the optical path, we consider transmission through three distinct regions: the input path just before the cryostat; the path through the cryostat; and the output path from the cryostat to the OSA. We directly measure the power incident to the cryostat via a calibrated 99:1 beam splitter. To measure the insertion loss through the cryostat, we optimize the optical polarization in the vicinity of the three modes we care about. However, we cannot disambiguate the insertion loss of the input grating coupler and the output grating coupler; so the best we can do is assume the loss is equal. For the measurements in this work, we find the single-sided grating coupler efficiency to be $\eta_\text{input}^\text{opt} = 8.84 \%$. Lastly, the optical path loss between the cryostat output and the OSA is $82.3 \%$. This gives a total output efficiency of $\eta_\text{output}^\text{opt} = 7.28 \%$. 

\begin{figure}[h!]
    \centering
    \includegraphics[width=\linewidth]{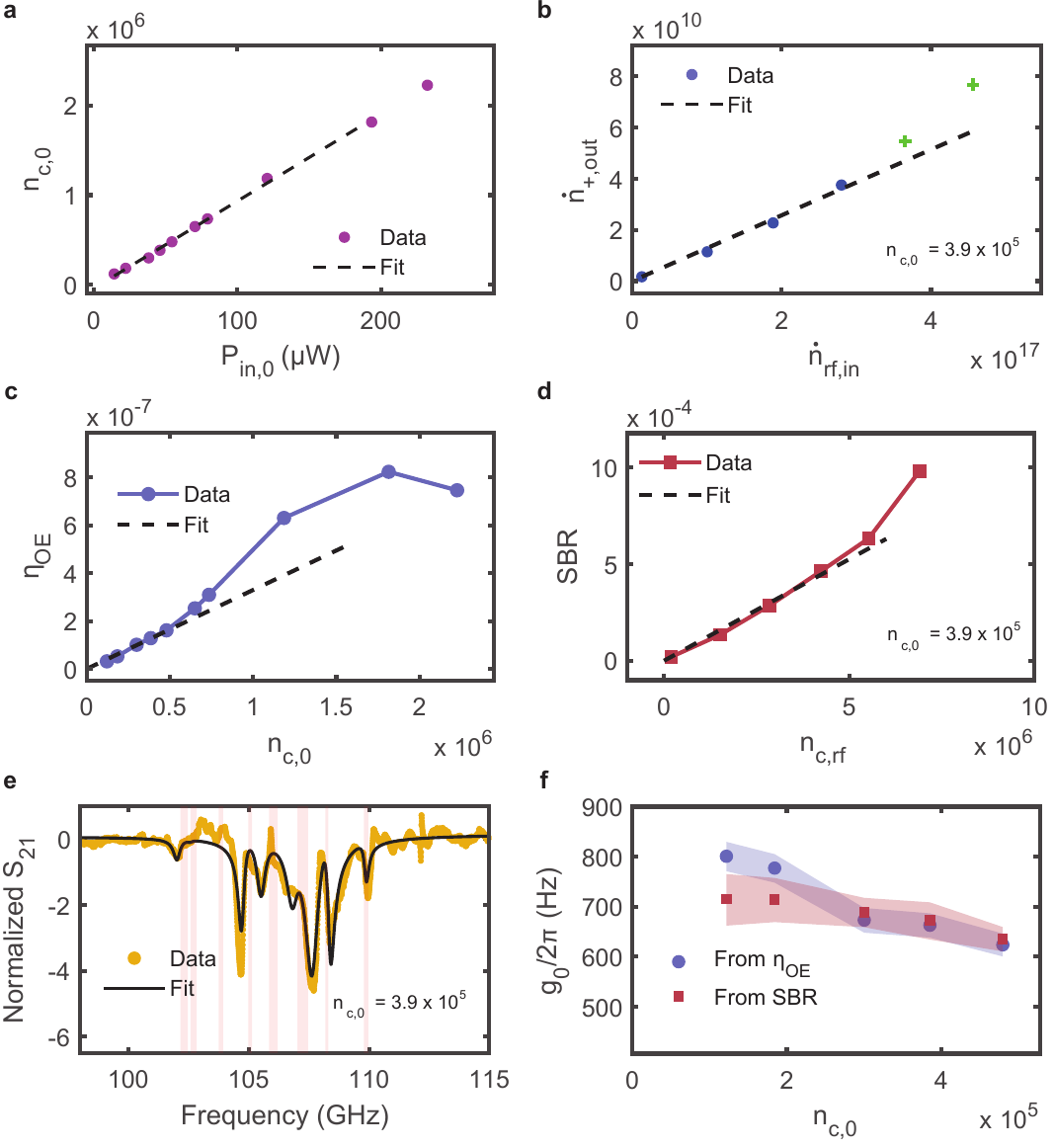}
    \caption{Step-by-step data analysis required to extract the electro-optic coupling. \textbf{a} A calibrated plot of the intracavity pump photon number versus on-chip optical pump power. The data is nearly linear, with variation coming from differences in the lock-point detuning of the pump tone from resonance. \textbf{b} A representative example of sideband photon flux plotted against RF photon flux. At low RF powers, this ratio is linear and can be fit to determine $\eta_\mathrm{OE}$. \textbf{c} Reproduction of the main text Fig. 3b. Each datapoint is determined by fitting $\eta_\mathrm{OE}$ as in (b). \textbf{d} Reproduction of main text Fig. 3c. \textbf{e} Representative example of RF spectrum and fit at a particular optical pump power. The red vertical bars are centered on substrate mode frequencies with widths matching the 3-dB linewidths of the modes. \textbf{f} Reproduction of main text Fig. 3e.}
    \label{fig:sfig1}
\end{figure}

%%%%%%%%%%%%%%%%%%%%%%%%%%%%%%%%%%%%%%%% SUBSECTION
\subsection{Determining the intracavity pump photon number}
\label{sec:optPhotOcc}
We plot the intracavity pump photon number as a function of optical power (in the waveguide) in~\hyperref[fig:sfig1]{Supplementary Figure 1a}. 
The ``y-axis'' of this plot depends on the precise detuning of our pump tone from the pump mode resonance frequency, $\omega_\mathrm{0}$. We determine this detuning for each input pump power by matching our measured transmission through the cavity with the theoretical transmission.
As described in the main text methods section, the following relation holds for the transmission through the optical resonator,
\begin{align}
    T &= \frac{P_\mathrm{OSA}}{P_\mathrm{PM}}\cdot \eta_0  \\
    &= \left|1- \frac{\kappa_{e,0}}{i\Delta_0 + \kappa_0/2} \right|^2.
\end{align}

$P_\mathrm{OSA}$ and $P_\mathrm{PM}$ are the optical pump power measured on the OSA (at the end of the measurement chain) and on a power meter (before sending the laser tone into the cryostat), respectively, and $\eta_0$ encapsulates optical path losses and calibrations on both the input and output paths. We measure every value in the second equality through our self-heterodyne characterization (see main text), except for the pump detuning, $\Delta_0=\omega_0 - \omega_p$. By equating these two expressions, we can solve for the detuning $\Delta_0$. We then compute the intracavity pump photon number, $n_{c,0} = \kappa_{\text{e},0}/(\Delta_0^2 + \kappa_0^2/4)\cdot\frac{P_{0}}{\hbar\omega_p}$. We repeat this process for every datapoint in~\hyperref[fig:sfig1]{Supplementary Fig. 1a}, thereby calibrating the ``x-axis'' of main text Fig. 3b.

%%%%%%%%%%%%%%%%%%%%%%%%%%%%%%%%%%%%%%%% SUBSECTION
\subsection{Determining the electro-optic conversion efficiency, $\eta_{\mathrm{OE}}$}
In this section, we provide our methodology to determine the electrical-to-optical transduction efficiency. For each applied pump power (each datapoint in~\hyperref[fig:sfig1]{Supplementary Fig. 1a}), we vary the applied RF modulation power. For each applied RF power, we vary the RF modulation frequency. At each frequency, we measure an OSA trace, such as that given in main text Fig. 3a. Then, we determine the OSA trace that yields the greatest optical sideband power, which we assume to be the experimental condition that satisfies the frequency matching between the RF modulation and the optical pump-to-sideband-mode frequency spacing. From these data we estimate the on-chip sideband photon flux, given our calibrations (discussed above). We also compute incident RF photon flux given the applied RF power, thereby obtaining a list of $\dot{n}_+^\mathrm{out}$ v.s. $\dot{n}_\mathrm{RF}^\mathrm{in}$. An example of these data is plotted in ~\hyperref[fig:sfig1]{Supplementary Fig. 1b}. By fitting a line to the lower-power datapoints in~\hyperref[fig:sfig1]{Supplementary Fig. 1b}, we determine the electrical-to-optical conversion efficiency. Measuring in this way allows us to average over fluctuations or errors in the value, which might enter the measurement at individual datapoints. We repeat this process for every optical pump power we apply to the device. For each optical power, we then plot the efficiency, thereby obtaining main text Fig. 3b (re-plotted as \hyperref[fig:sfig1]{Supplementary Fig. 1c}). 

The pseudo-code/algorithm to reproduce~\hyperref[fig:sfig1]{Supplementary Fig. 1c} is as follows (indents denote for-loop level, as in Python):
\bigskip
\setlength{\fboxsep}{20pt}
\begin{enumerate}
\doublebox{
\begin{minipage}{0.8\linewidth}  
    \item For each optical pump power from $P_\mathrm{1}^\mathrm{opt}$ to $P_n^\mathrm{opt}$:
    \begin{enumerate}
        \item For each RF modulation power from $P_\mathrm{1}^\mathrm{RF}$ to $P_m^\mathrm{RF}$:
        \begin{enumerate}
            \item For each RF modulation frequency from $\Omega_1$ to $\Omega_l$:
            \begin{enumerate}
                \item Measure the optical sideband spectrum on the OSA (see Fig. 3a, in the main text).
            \end{enumerate}
            \item Identify the spectrum with the greatest optical sideband power and record both the pump and sideband powers.
            \item Infer the on-chip optical photon flux and RF photon flux using the path loss calibrations given in~\hyperref[calibrations]{Section S2.1}.
        \end{enumerate}     
        \item Plot and linearly fit $\dot{n}_+^\mathrm{out}$ v.s. $\dot{n}_\mathrm{RF}^\mathrm{in}$. The slope of this line is $\eta_\mathrm{OE}$.
    \end{enumerate}
    \end{minipage}
}
\end{enumerate}
\vspace{10pt}

%%%%%%%%%%%%%%%%%%%%%%%%%%%%%%%%%%%%%%%% SUBSECTION
\subsection{The electro-optic coupling rate, $g_0$ as a function of optical power}
Here we explain how we obtain an estimate for the electro-optic coupling rate as a function of optical power. In principle, these two quantities are independent of one another. However, since the sub-THz cavity parameters (i.e., frequency and linewidth) depend on optical power, we can get multiple estimates of $g_0$ (one for each applied optical power). In Fig. 3b of the main text, reproduced as~\hyperref[fig:sfig1]{Supplementary Fig. 1c}, we see that the first five datapoints in this plot grow linearly with intracavity photon number, as expected from theory. We obtain the slope of this linear region and, by incorporating the optical power dependence of our sub-THz cavity mode (Equation S35), we infer five separate values of $g_0$ (see \hyperref[fig:sfig1]{Supplementary Fig. 1f}). More details on the fitting procedure to obtain the sub-THz cavity parameters is found in~\hyperref[substratefit]{Section S3}.

%%%%%%%%%%%%%%%%%%%%%%%%%%%%%%%%%%%%%%%% SUBSECTION
\subsection{Determining the intracavity RF photon number}
\label{sec:rfPhotOcc}
The RF photon cavity occupation is a function of temperature, optical power, and its coupling to the surrounding substrate modes. This quantity is directly used in inferring the $g_0$ from the sideband ratio, $\Upsilon_{+,0}$ (see Eq. S36 and section below). Experimentally, we measure the RF spectrum for each optical pump power and RF power. We fit the low-RF-power trace at each optical power, taking into account the various substrate modes. The linewidths and frequencies of these substrate modes are previously determined from fitting the RF spectrum at various cryostat temperatures (as explained in the main text and in~\hyperref[substratefit]{Supplementary Section S3}). Therefore, with the substrate mode parameters fixed, each fit determines only the superconducting sub-THz resonance frequency and linewidth at that particular applied optical power. In~\hyperref[fig:sfig1]{Supplementary Fig. 1e} we depict one such trace and fit; the center frequency and width of the red vertical bars denote the substrate mode frequencies and linewidths, respectively. The mode parameters from the fit are then used to determine the intracavity RF population using Eq. S31.  

%%%%%%%%%%%%%%%%%%%%%%%%%%%%%%%%%%%%%%%% SUBSECTION
\subsection{The sideband ratio as a function of optical pump power}
As with the electro-optic coupling rate, the sideband ratio (SBR) is, in principle, uncorrelated with the optical pump power. In practice, however, the SBR depends on optical pump power due to the optical power dependence of the sub-THz cavity parameters. The sideband ratio (SBR) is defined as the ratio between the transmitted power in the blue sideband frequency versus the transmitted power at the optical pump frequency. For a given optical pump power, we compute the SBR for each of the applied RF modulation powers. A reproduction of this plot from main text Fig. 3c is given in~\hyperref[fig:sfig1]{Supplementary Fig. 1d}. We fit a line to the linear regime of this plot (corresponding to low RF modulation powers). This line has a slope proportional to $g_0^2$ (see Eq. S36). Similar to $\eta_\mathrm{OE}$ (see above), we solve for $g_0$ from this slope, accounting for the shifted superconducting RF mode and substrate modes separately for each datapoint.  A reproduction of main text Fig. 3e is given in~\hyperref[fig:sfig1]{Supplementary Fig. 1f}. This figure presents the values of $g_0$ determined from both the SBR and $\eta_\mathrm{OE}$-fitting methods.
\clearpage

%%%%%%%%%%%%%%%%%%%%%%%%%%%%%%%%%%%%%%%% SECTION
\section{Methodology for fitting the multiparameter substrate mode model}
\label{substratefit}
In our experiment, we observe dips in the RF spectrum that we associate with substrate modes. These are high-frequency electromagnetic modes that are resonant in the approximately $\qty{500}{\um}$-thick Sa substrate of our device. The presence of these modes is evidenced by simulations (see Methods), and we derive a model describing their interaction with the superconducting (SC) sub-THz mode in \hyperref[substratemodes]{Supplementary Section S1.3}. Importantly, if there is overlap between the electromagnetic field of the mode in the SC resonator and the field of the substrate mode, the two modes will couple, leading to an effective detuning imparted on the SC resonator mode frequency, and a broadening of its linewidth (inducing an effective loss of sub-THz photons from the SC resonator). Our model (see Eqs. S29-S34) and our fitting of this model to our data, adheres to the following assumptions:
\vspace{10pt}
\setlength{\fboxsep}{20pt}
\begin{enumerate}
    \doublebox{
        \begin{minipage}{0.8\linewidth}
            \item The substrate modes are \textit{only} excited through hybridization with SC resonator mode; they cannot be driven directly by the impinging RF field inside the WR10 waveguide.
            \item Both the total loss rate ($\kappa_\mathrm{RF}$) and the coupling loss rate ($\kappa_{e\mathrm{,RF}}$) of the SC mode can only increase as temperature increases.    
            \item The SC resonator frequency can only decrease (``red-shift'') with increasing temperature.
        \end{minipage}
    }
\end{enumerate}    

\vspace{10pt}
Note that the temperature can be increased by both changing the temperature inside the cryostat thermally, or by pumping enough light into the optical resonator to locally heat the SC resonator. In order to determine the temperature dependence and parameters of the SC sub-THz mode and the substrate modes, we independently sweep the cryostat temperature (with optical power off) and record the RF spectra, and sweep the incident optical power to the chip (with the cryostat at base temperature), and record the RF spectra. We then fit the recorded spectra as described in the following sections. We also tested the results of fitting with a modified assumption ``2'' restricting the coupling loss rate, $\kappa_{e\mathrm{,RF}}$ to \textit{only decrease} (while the total loss rate is still restricted to only increase). In this case, we found our predictions for $\eta_\mathrm{OE}$ and $g_0$ do not vary much from the reported values and are within the reported confidence intervals. Therefore, we report the results in this manuscript using the original assumption ``2'' given above. 

%%%%%%%%%%%%%%%%%%%%%%%%%%%%%%%%%%%%%%%% SUBSECTION
\subsection{Temperature-dependent fitting procedure}
Using particle swarm optimization (PSO), we iteratively fit the temperature-swept RF data to infer the parameters of each substrate mode. We first identify eight substrate modes that couple to the SC resonance as it thermally tunes. Our model for RF transmission, $S_{21}^{\mathrm{RF}}$ is then given by equation S34. We also include a linear background that is allowed to vary for each spectrum, yielding an expression comprising 29 free parameters: the SC mode parameters $\kappa_\mathrm{RF}$, $\kappa_{e,\mathrm{RF}}$, $\Delta_\mathrm{RF}$; the substrate mode parameters $\gamma_n$, $\omega_n$, $J_n$ for $n\in\{1,2,...,8\}$; and the linear background slope $m$ and intercept $b$. The full expression is then:
\begin{equation}
    S_{21}^{\mathrm{RF}} = \left(1- \frac{\kappa_{e,\text{RF}}/2}{i(\Delta_\text{RF}+\tilde \delta) + (\kappa_\text{RF} + \tilde\gamma)/2}\right) + m\Omega + b
\end{equation}
where $\tilde\delta$ and $\tilde\gamma$ are given by equations S32 and S33 respectively, with $\Delta_n = \omega_n-\Omega$.

We repeatedly fit the temperature-dependent spectra and average over the results to fix additional parameters of these modes, thereby reducing the number of parameters that subsequent iterations of the PSO must optimize. Our detailed procedure follows:
\vspace{10pt}
\setlength{\fboxsep}{20pt}
\begin{enumerate}
    \doublebox{
      \begin{minipage}{0.8\linewidth}
            \item Fit four RF spectra at the temperatures $4.89$ K, $6.00$ K, $6.70$ K, and $7.01$ K, normalized to a high-temperature background spectrum (at $13.01$ K). Identify eight common substrate modes across these spectra.            
            \item Iteratively fit  $|S_{21}^{\mathrm{RF}}|^2$ over $22$ temperatures ranging from $4.89$ K to $6.90$ K, allowing all parameters to vary. After fitting all $22$ spectra, we average over the fit results for the frequencies of the substrate modes, $\omega_{n}$. We only include fit results in the averaging for which the fit of that particular substrate mode looks reasonable. An example comparing such fits is given in~\hyperref[fig:sfig2]{Supplementary Fig. 2}.
            \item Fix some of the frequencies and repeat step (2) until all $\omega_n$ are fixed. 
            \item Repeat the procedure in steps (2)-(3), but now with fixed frequencies, $\omega_n$ and allowing only the coupling rates $J_n$ and loss rates $\gamma_n$ to vary. As before, average over all reasonable fits to fix the $\gamma_n$ of each mode.
            \item Lastly, fix both $\omega_n$ and $\gamma_n$ of all substrate modes. Fit all $22$ temperature spectra, allowing only $J_n$, the linear background, and the SC mode parameters to vary, thereby obtaining how $\omega_\mathrm{RF}$ and $\kappa_\mathrm{RF}$ tune with temperature.
        \end{minipage}
    }
\end{enumerate}
\vspace{10pt}

Examples of fits from this process are given in \hyperref[fig:sfig2]{Supplementary Fig. 2}. These examples demonstrate which types of fits would be included in the average of the frequency (or comparably the loss rate) for a given substrate mode.

\begin{figure}[h!]
    \centering
    \includegraphics[width=\linewidth]{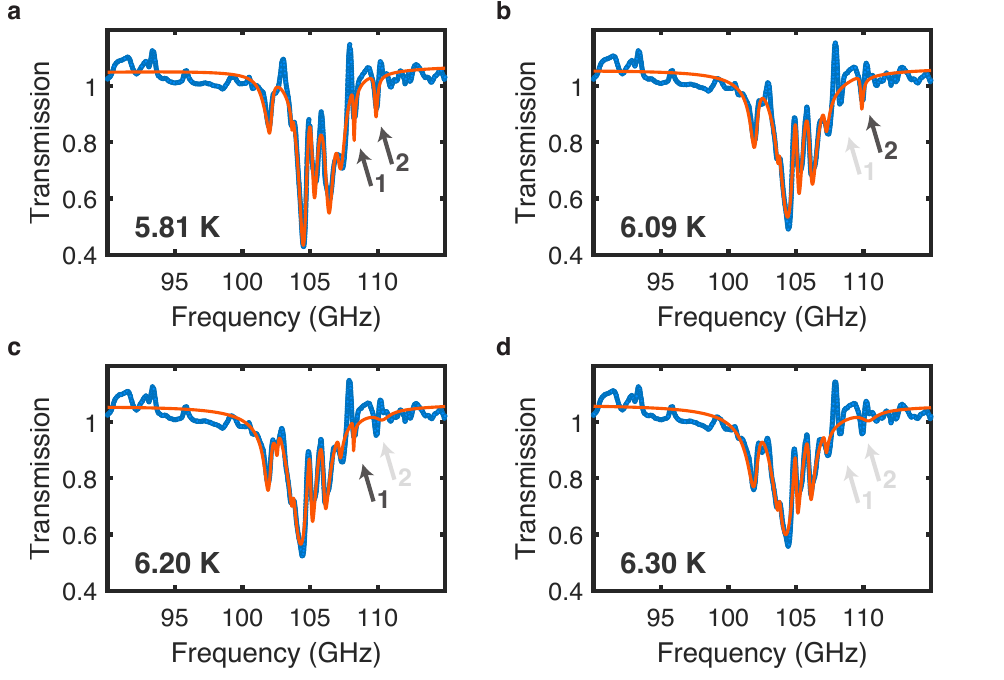}
    \caption{RF spectrum fits at four different temperatures. Transmission is normalized to a high-temperature (non-superconducting) background spectrum. Blue is data, and red is the resulting fit from PSO optimization. Dark grey arrows indicate two example substrate modes ``1'' and ``2.'' \textbf{a} RF spectrum taken at platform temperature $\sim5.81$ K. The dark grey arrows indicate that the fit results from this spectrum would be included in the average to determine the frequencies of both modes ``1'' and ``2.'' \textbf{b} RF spectrum taken at platform temperature $\sim6.09$ K. The dip corresponding to substrate mode ``1'' is missing from the fit, so this result would not be included in the average to determine the frequency of mode ``1.'' However, the dip corresponding to mode ``2'' is still reasonable, so we would include this result when averaging for the frequency of mode ``2.'' \textbf{c} RF spectrum taken at platform temperature $\sim6.20$ K. This is the opposite situation to (b). The fit for mode ``1'' is reasonable, while the fit for mode ``2'' is poor. We would only include this result in the average to determine the frequency of mode ``1,'' but not for the frequency of ``2.'' \textbf{d} RF spectrum taken at platform temperature $\sim6.30$ K. In this case neither dip is present, so we would not include the results of this fit in the averaging to determine either mode ``1'' or ``2.''}
    \label{fig:sfig2}
\end{figure}

%%%%%%%%%%%%%%%%%%%%%%%%%%%%%%%%%%%%%%%% SUBSECTION
\subsection{Optical-power-dependent fitting procedure}
After identifying $\omega_n$ and $\gamma_n$ for all substrate modes, we repeat step (5) of the above procedure, but over optical-power-dependent spectra at a fixed RF power. Using only low-RF-power datasets, we allow the SC resonance parameters to vary, along with the coupling rate, $J_n$ ($n\in [1,\ldots,8]$) of the eight substrate modes, and the linear background of each spectrum. From these fits, we identify how the SC mode frequency and loss rate tune with increasing optical power.
\clearpage

%%%%%%%%%%%%%%%%%%%%%%%%%%%%%%%%%%%%%%%% SECTION
\section{Additional details on experimental setup}
\label{experimentalsetup}

%%%%%%%%%%%%%%%%%%%%%%%%%%%%%%%%%%%%%%%% SUBSECTION
\subsection{Primary optical path and wavelength calibration}
\label{sec:opticalpath}
The optical path used for characterizing the on-chip sub-THz modulation and optical sideband generation is shown in~\hyperref[fig:sfig3]{Supplementary Figure 3}. Light from a telecom laser (Santec TSL-710) is first attenuated via a MEMs variable optical attenuator (VOA). It then passes through a beam splitter, with most of the light continuing to the cryostat (Montana Instruments S200 Cryostation). A fraction of the light is passed into a wavelength calibration sub-path.

In the calibration sub-path, the light is split with a $50:50$ beamsplitter, with half passed to a wavemeter (WM, Bristol 621 Wavelength Meter) for wavelength measurement, and the other passed through a Mach-Zehnder interferometer (MZI), with a measured FSR of $\sim325$ MHz). While collecting optical spectra of our device, we record the initial wavelength of the laser on the WM and then sweep the laser, simultaneously recording transmission through the device (DUT) and the fringing from the MZI. By combining the WM reading with the measured MZI spectrum, we can calibrate the wavelength at each point in the recorded device spectrum. While there is some error in the exact wavelength reported by the WM, the calibrated wavelength spacing between recorded data points is much more accurate, ensuring confident measurements of wavelength \textit{ranges}, such as the optical mode linewidths and the free spectral range (FSR) of the racetrack resonator.

In the primary path, light passes through a circulator (to suppress reflections), followed by a 2x2 ``x-switch'' (Sercalo, SL2x2). With the switch in a bar configuration, the optical signal bypasses the EOM self-heterodyne sub-path (see \hyperref[sec:EOM_heterodyne_subpath]{Section S4.2}), continues through another beamsplitter (used for optical power calibration), and is incident to the DUT inside the cryostat. The cryostat is maintained at nearly $5$ K for the duration of measurements in this paper. The output light from the DUT, exiting the cryostat, is split into two arms. A fraction of the signal is incident upon an avalanche photodiode (APD) for broadband optical spectrum recording. The majority of the signal is passed to an optical spectrum analyzer (OSA, Yokogawa AQ6374) for characterizing the wavelengths and powers of generated sidebands when RF modulation is turned on. We measure the insertion losses and splitting ratios of the beam splitters immediately before and after the cryostat/DUT, thereby estimating losses induced by the cryostat optical ports and the couplings between the glued optical fibers and the DUT.

\begin{figure*}[b]
    \centering
    \includegraphics[width=\linewidth]{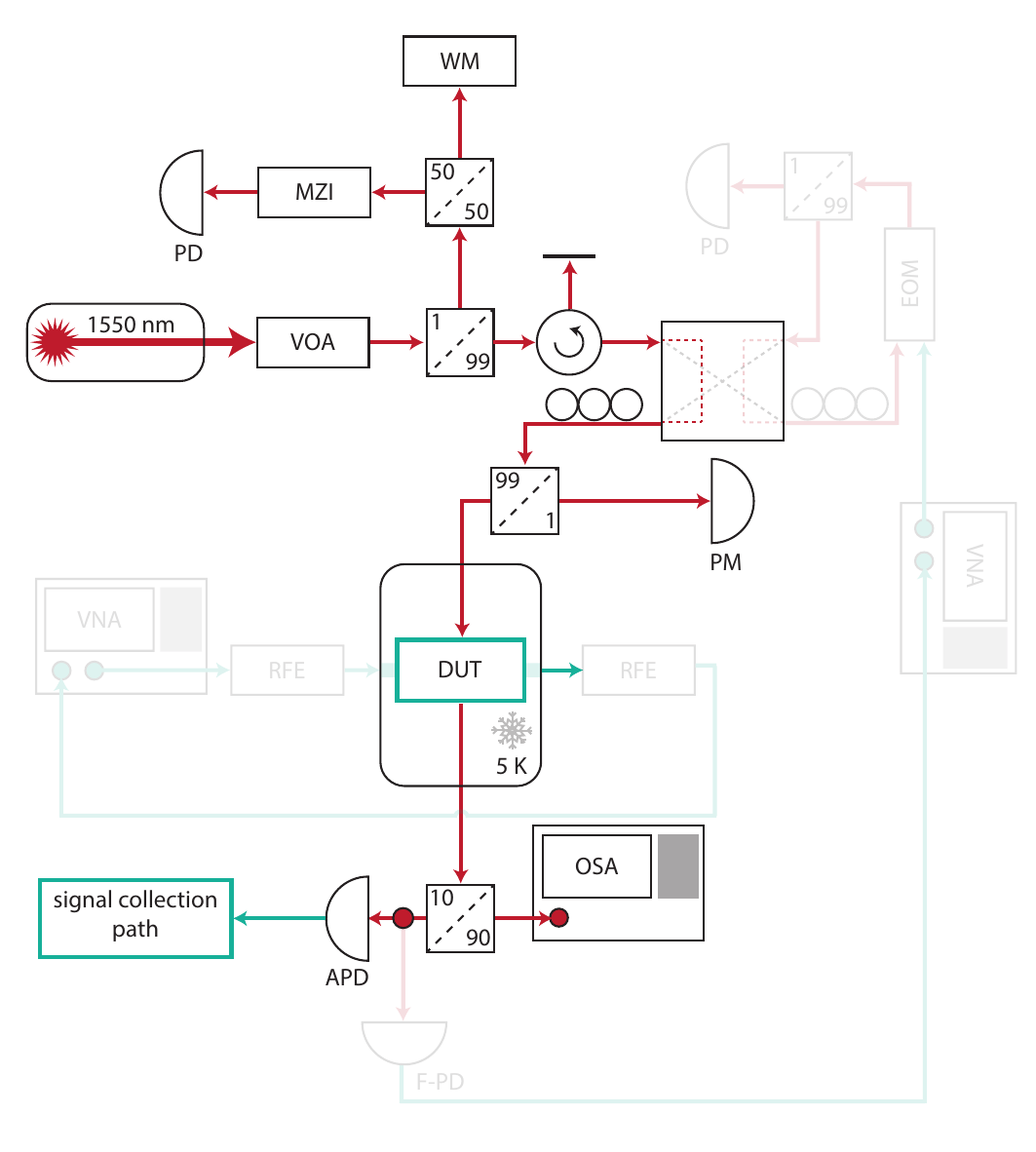}
    \caption{Primary optical path.}
    \label{fig:sfig3}
\end{figure*}

\subsection{Self-heterodyne measurement path}\label{sec:EOM_heterodyne_subpath}
We characterize the optical modes, $\omega_k$, independently using a self-heterodyne measurement, similar to those presented by Patel et al.~\cite{patel:2018:singlemodephononic} and Herrmann et al.~\cite{herrmann:2022:mirrorsymmetric}. The path for this measurement is depicted in~\hyperref[fig:sfig4]{Supplementary Figure 4}.

For each optical mode, we feed back on the WM reading to lock the incident laser blue-detuned from resonance by a few GHz. With the Sercalo SL2x2 switch in cross configuration, light passes through a separate off-chip electro-optic intensity modulator. We modulate the EOM with the output from a VNA (Rhode \& Schwartz ZNB), thereby generating sidebands on the locked incident laser. We pass the pump, along with the sidebands, to the DUT. By sweeping the modulation frequency on the VNA, the red sideband from the EOM sweeps across the optical resonance. We detect the output pump light and both sidebands on a fast photodiode (F-PD, Optilab PD-40-M) and pass the resulting electrical signal, consisting of beat tones between the locked pump and swept sidebands, on the receiving port of the VNA. We normalize this spectrum to a VNA trace measured with the incident pump very far-detuned from resonance. Finally, we fit the result to an input-output model, from which we infer the parameters of each optical mode. We present the results of this procedure applied to the optical pump mode $\omega_0$, in main text Fig. 2b,c.

\begin{figure*}[h]
    \centering
    \includegraphics[width=\linewidth]{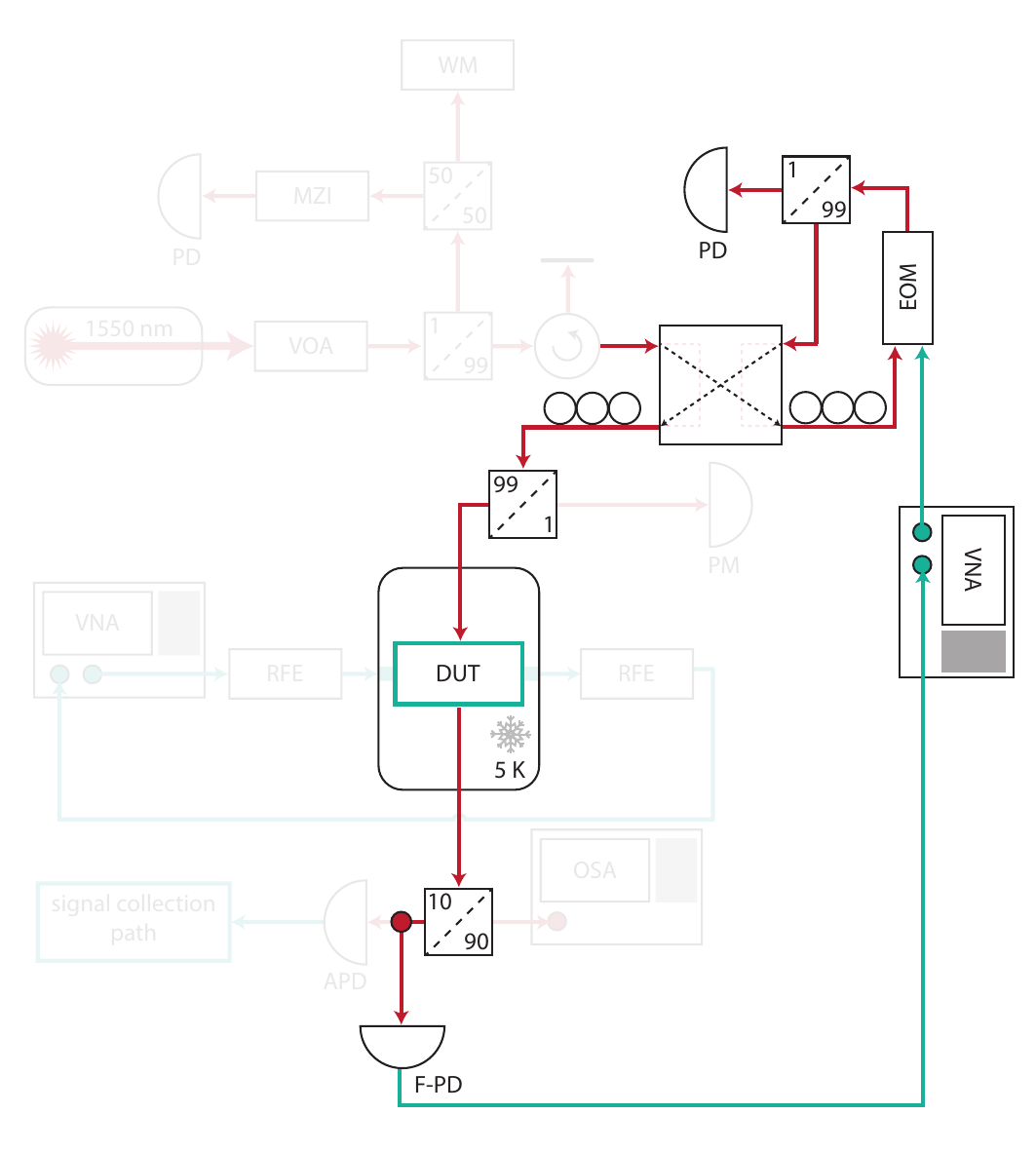}
    \caption{Self-heterodyne sub-path for optical mode characterization.}
    \label{fig:sfig4}
\end{figure*}

\subsubsection{Self-heterodyne measurement theory}
\label{opticalphase}
In this section, we sketch the theoretical basis for the phase response measurement we performed to measure the optical cavity parameters discussed in the previous section. Consider the forward scattering matrix of a lossless, symmetric, and reciprocal Mach-Zehnder interferometer (MZI)~\cite{patel:2018:singlemodephononic, bucholtz:2020:matrixrepresentations, herrmann:2022:mirrorsymmetric},

\begin{align}
    \begin{pmatrix}
\alpha_\text{in} \\
\cdot \\
\end{pmatrix}
 &=
\Big(\frac{1}{\sqrt{2}}\Big)^2\cdot
\begin{pmatrix}
1 & i \\
i & 1
\end{pmatrix}
\begin{pmatrix}
1 & 0 \\
0 & e^{i\phi}
\end{pmatrix}
\begin{pmatrix}
1 & i \\
i & 1
\end{pmatrix}\cdot
\begin{pmatrix}
\alpha_\text{L} \\
0
\end{pmatrix} \implies \\
\alpha_\text{in} &= \frac{1}{2}\cdot\Big(1 - e^{i\phi} \Big) \cdot \alpha_\text{L}.
\end{align}
Here $\phi$ is the relative phase delay in one arm of the MZI, $\alpha_\text{L}$ is the amplitude of the laser field, and $\alpha_\text{in}$ is the field we send to our optical cavity in the frame of the laser. In our case, we can vary the phase with a sinusoidal voltage at frequency $\Omega$, so that $ \phi = \phi(t) = \theta_\text{DC} + \beta\cos(\Omega t)$. This voltage signal comes from the source port of the VNA and is combined at the modulator with a DC voltage. Operating the modulator at mid-point so that $\theta_\text{DC} = \pi/2$ and assuming small modulation strength $\beta$ we calculate,
\begin{align}
    \alpha_\text{in}(t) = \frac{1}{2}\cdot\Big((1-i) + \frac{\beta}{2}e^{-i\Omega t} + \frac{\beta}{2}e^{i\Omega t}\Big)\cdot \alpha_\text{L}.
\end{align}
We can write the frequency response of the optical cavity transmission in the frame of the source laser as,
\begin{align}
    \mathcal{T}(\omega; \Delta) = 1 - \frac{\kappa_e}{i(\Delta-\omega) + \kappa/2},
\end{align}
where $\Delta = \omega_c-\omega_\text{L}$, the detuning between the optical cavity mode frequency and the source laser frequency. Because the optical cavity is a linear time-invariant system we can write the transmitted field amplitude as (omitting $2\Omega$ terms),

\begin{align}
    \alpha_\text{trans}(t;\Delta) &= \frac{\alpha_\text{L}}{2} \cdot \Big((1-i)\mathcal{T}(0; \Delta) + \frac{\beta}{2}e^{-i\Omega t}\mathcal{T}(\Omega; \Delta) + \frac{\beta}{2}e^{i\Omega t}\mathcal{T}(-\Omega; \Delta) \Big).
\end{align}
The transmitted field is routed to a fast photodetector so that the beat frequency generates an RF signal that is AC-coupled to the receiving port of the VNA. Thus, the AC-coupled field is proportional to,

\begin{align}
    |\alpha_\text{trans}(t;\Delta)|^2 &= \frac{|\alpha_\text{L}|^2\beta}{8} \cdot \Big((1+i)\mathcal{T}^\ast(0; \Delta)\mathcal{T}(\Omega) +  (1-i)\mathcal{T}(0; \Delta)\mathcal{T}^\ast(-\Omega; \Delta)\Big)\cdot e^{-i\Omega t} \notag\\
    &+ \text{c.c}. 
\end{align}
On the VNA, we measure the Fourier transform of this expression, equivalent to the RF scattering parameter $S_{21}(\Omega)$ (i.e. the coefficient of the exponential). However, because this expression is proportional to the laser power ($|\alpha_\text{L}|^2$) and the RF power ($\beta$), this measurement includes optical path losses and the frequency responses of the RF circuit elements. We normalize out these frequency-dependent losses by taking another measurement with the source laser detuned far away from the mode so that $\Delta \gg \kappa$. In this case, $\mathcal{T} \approx 1$ and we obtain:
\begin{align}
\Big|\alpha_\text{trans}^\text{bg}\Big|^2 &= \frac{|\alpha_\text{L}|^2\beta}{4}\cdot e^{-i\Omega t} + \text{c.c.}
\end{align}
Dividing the signal response by the far-detuned background, we obtain a normalized expression for the RF scattering parameter,
\begin{align}
S_{21}^\text{normalized}(\Omega; \Delta) &= \frac{1}{2} \Big((1+i)\mathcal{T}^\ast(0; \Delta)\mathcal{T}(\Omega) +  (1-i)\mathcal{T}(0; \Delta)\mathcal{T}^\ast(-\Omega; \Delta)\Big).
\end{align}
We use Eq. S50 and fit the phase response ($\angle S_{21}^\text{normalized}$) to deduce all the optical cavity parameters.

\subsection{Primary RF path}
The primary RF modulation path in our experiment. The output signal from a specialized VNA (Rhode \& Schwartz ZNA26) is up-converted through frequency mixing in a radio-frequency extender (RFE, OML Inc.) as highlighted in~\hyperref[fig:sfig5]{Supplementary Fig. S5}. The RFE effectively translates the VNA range from $0-26$ GHz into $70-115$ GHz.

The RFE output is transitioned from WR10 waveguide into a $1$ mm coaxial cable, which is fed into the cryostat via a hermetically sealed bulkhead connector. Inside the cryostat, the coaxial cable is translated back into a WR10 waveguide, enabling the RF signal to couple capacitively to the chip as explained in the main text. Transmission through the chip passes through an W-Band cryogenic isolator (Micro Harmonics), a W-Band cryogenic HEMT amplifier (Low Noise Factory LNF-LNC65\_115WB), and another isolator (Micro Harmonics), before coupling back into $1$ mm coaxial cable and routed out of the cryostat. This signal is down-converted to the native VNA range with a second RFE and recorded on the receiving port of the VNA. 

\begin{figure*}[h]
    \centering
    \includegraphics[width=\linewidth]{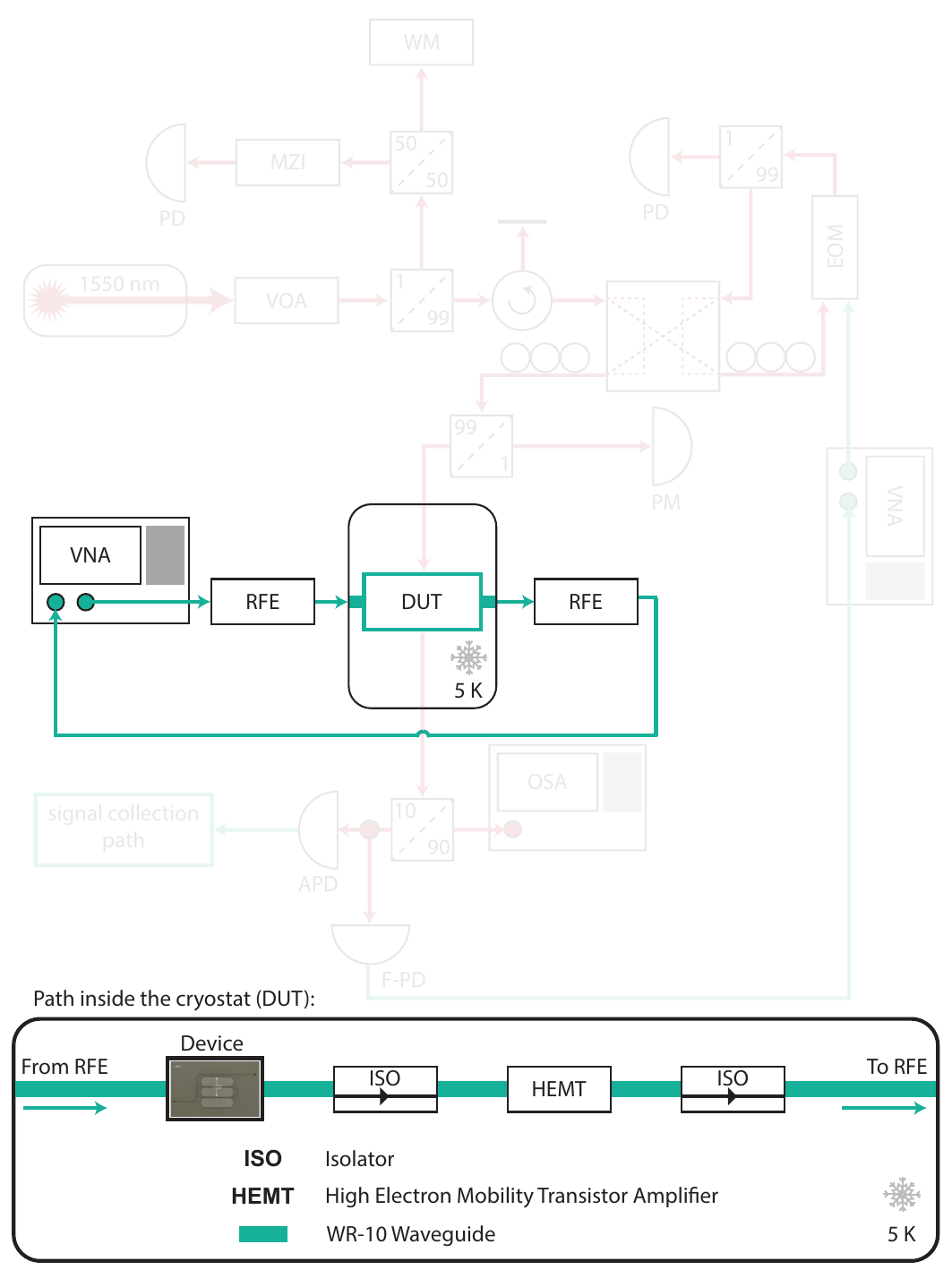}
    \caption{Primary RF path for on-chip sub-THz modulation. RFEs are used to up-/down-mix the VNA output into the range of $70-115$ GHz. The signal passes through the device in transmission. The transmission passes through an isolator, a HEMT amplifier, and another isolator, before exiting the cryostat and being collected/recorded on the VNA. }
    \label{fig:sfig5}
\end{figure*}

\clearpage
\subsection{Optical-pump locking electronics path}
\label{sec:pumplock}
The electronic sub-path used for optical pump locking is shown in~\hyperref[fig:sfig6]{Supplementary Figure 6}. Before each transduction measurement, we must lock the incident laser in the pump mode $\omega_0$, and calibrate its detuning from resonance. The mode drifts under locking due to thermal and photorefractive effects, so we lock slightly blue-detuned of the resonant frequency for increased stability. 
\begin{figure*}[hb]
    \centering
    \includegraphics[width=\linewidth]{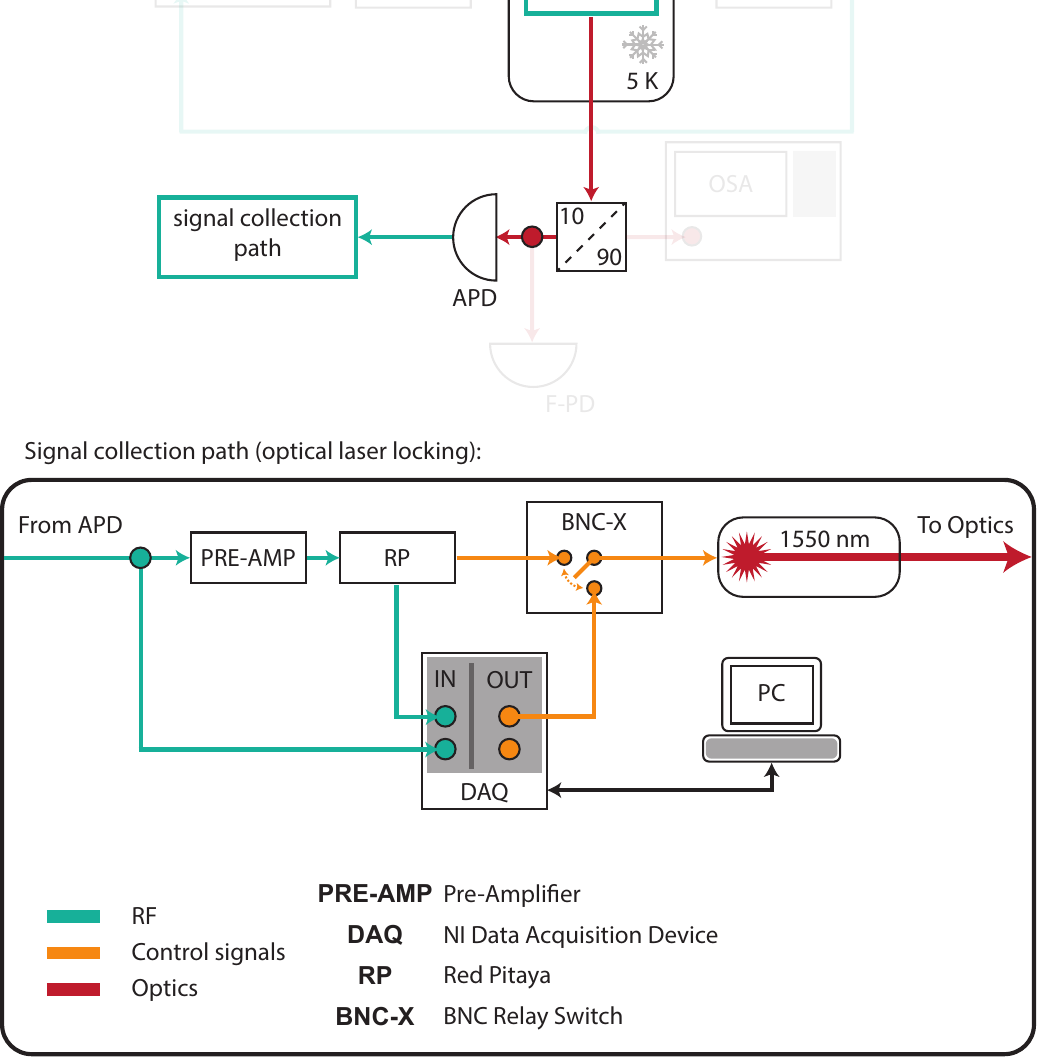}
    \caption{Electronic signal collection sub-path. This path is used for locking the incident pump laser to the optical mode, $\omega_0$. The output of the APD is split into two paths, one recorded directly on a DAQ and PC, and the other passed through a pre-amp to the Red Pitaya voltage input. The output of the Red Pitaya and the DAQ are both used to modulate the laser, and a BNC mechanical relay switch is used to swap which output controls the laser.}
    \label{fig:sfig6}
\end{figure*}
\clearpage

%%%%%%%%%%%%%%%%%%%%%%%%%%%%%%%%%%%%%%%% SECTION
\section{Cryogenic photorefractive behavior and in-situ mitigation}
 During our first locking attempts at base temperature ($\sim$4-5 K) we observed strong photorefractive effects, not uncommon in oxide-clad TFLN resonators~\cite{xu:2021:mitigatingphotorefractive}. Photo-excited charge carriers (from the optical pump) can distribute through the crystal, setting up a space-charge field that causes a refractive index change via the electro-optic effect. The practical impact of this in our measurements is that it limits how long our pump laser can remain locked to the optical pump mode. The maximum lock time is given by  $t_\text{lock} = (1 \text{ V}) \cdot (dV_\text{piezo}/dt)$, where 1 V is the maximum analog output from our Red Pitaya (0 V to 1 V). 

\begin{figure}[h!]
    \centering
    \includegraphics[width=1\linewidth]{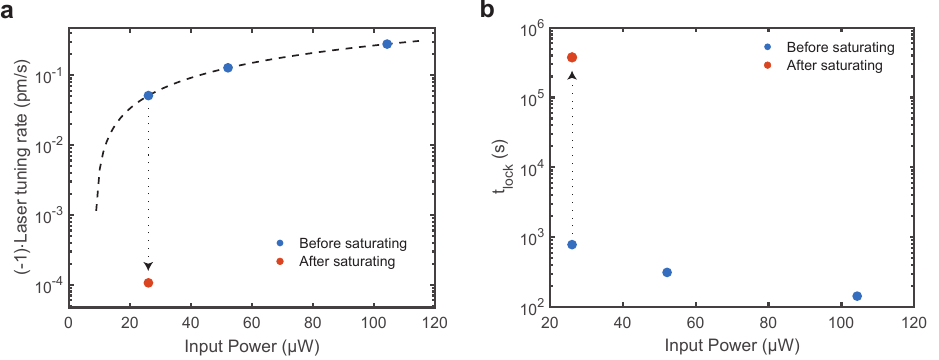}
    \caption{Drift due to photorefractive effect. \textbf{a} In this semi-log plot, we show the laser tuning rate (note, that this value is actually negative) as a function of optical power in the TFLN waveguide. The blue data show the drift rates before we reach steady-state (saturation). Near saturation, the drift rate reduces roughly 3 orders of magnitude, indicated by the dashed arrow and red data. The dashed line indicates a linear fit $y = (-2.9 \cdot 10^3 \text{ pm/W/s}) \cdot P_\text{in} + 0.03 \text{ pm/s}$. \textbf{b} The corresponding data converted to the maximum lock time. }
    \label{fig:sfig7}
\end{figure}

 To quantify some of these effects, we measure the drift rate directly by looking at how fast the piezo-voltage set point changes from the Red Pitaya, while we are locked on a particular optical mode (see Sec.~\hyperref[sec:pumplock]{S4.4}). The blue data in \hyperref[fig:sfig7]{Supplementary Fig. 7a} shows the laser tuning rate, ($dV_\text{piezo}/dt) \cdot(-40 \text{ pm/V})$, as a function of optical power. We can interpret this tuning rate as being equal to the cavity tuning rate, because we are locked at a fixed detuning from the cavity mode. The corresponding maximum lock time is shown in \hyperref[fig:sfig7]{Supplementary Fig. 7b}.

 To saturate the effect, we lock to an optical mode with a power low enough that the lock time is on the order of 10 minutes. We repeat this until enough carriers are excited to stabilize the space-charge field.  At this point, the lock time is much longer than we require to perform our electro-optic characterization. This procedure is indicated by the dashed arrow in \hyperref[fig:sfig7]{Supplementary Fig. 7}.

\end{document}